\documentclass[twocolumn, 
        showkeys, 
        jpb,
        superscriptaddress,
        floatfix,
        nofootinbib,pre]{revtex4-2}
\usepackage{graphicx} 

\usepackage[a4paper, top=2cm, bottom=2cm, left=2cm, right=2cm]{geometry}
\usepackage{amsmath}
\usepackage{xcolor}
\usepackage{comment}
\usepackage[normalem]{ulem}
\usepackage{hyperref}
\usepackage{amssymb} 
\usepackage{physics}

\begin{document}
\title{Passive memory reshapes active persistence}

\author{Ivan Di Terlizzi}
\email{These authors contributed equally to this work}
\affiliation{Max Planck Institute for the Physics of Complex Systems, Nöthnitzer Straße 38, 01187 Dresden, Germany}
\affiliation{Ludwig-Maximilians-Universit\"at M\"unchen, Arnold-Sommerfeld-Center for Theoretical Physics, M\"unchen, Germany}
\author{Lara Koehler}
\email{These authors contributed equally to this work}
\affiliation{Max Planck Institute for the Physics of Complex Systems, Nöthnitzer Straße 38, 01187 Dresden, Germany}
\affiliation{Cluster of Excellence Physics of Life, TU Dresden, Dresden, 01307 Germany}
\author{John D. Treado}
\email{These authors contributed equally to this work}
\affiliation{Max Planck Institute for the Physics of Complex Systems, Nöthnitzer Straße 38, 01187 Dresden, Germany}
\affiliation{Cluster of Excellence Physics of Life, TU Dresden, Dresden, 01307 Germany}

\bibliographystyle{unsrt}  

\begin{abstract}
Many active systems move in complex environments whose mechanical response is slow and history dependent. To address this regime, we study the collective dynamics of self-sustained active particles in non-Markovian media within a generalized Langevin framework with memory. We focus on the competition between the timescales of active persistence and viscoelastic relaxation in the environment. Using a minimal interacting model with an exponential memory kernel, we show that environmental memory qualitatively reshapes motility-induced phase separation of self-propelled active particles. When the memory timescale is comparable to the active persistence time, delayed response generates an effective anti-persistence that suppresses clustering and produces a broad metastable regime with slow nucleation dynamics. By contrast, for long memory timescales, reduced friction at short times enhances the effective propulsion velocity and restores phase separation. Our results demonstrate that the surrounding medium can actively regulate the emergence, stability, and dynamics of collective organization in active matter.
\end{abstract}

\maketitle

Active matter is often modeled in environments whose mechanical response relaxes rapidly compared with the persistence time of self-propulsion, allowing the surrounding medium to be treated as effectively Markovian \cite{fily2012athermal,zottl2023modeling}. 
Many natural and synthetic active systems, however, move in media with slow mechanical relaxation and long-lived memory effects \cite{bechinger2016active,molina2018crossover}, such as synthetic Janus colloids in polymeric solutions, where viscoelastic stresses feedback on particle orientation \cite{gomez2016dynamics,saad2019diffusiophoresis}; bacterial swimmers in mucus or polymer-rich environments, where elastic relaxation modifies persistence and angular dynamics \cite{patteson2015running,martinez2014flagellated,liao2023viscoelasticity, Liu.Wu.2021}; and motor-driven cargos moving through viscoelastic cytoplasm, where the medium retains deformations over times comparable to the stepping dynamics \cite{goychuk2014molecular,mickolajczyk2015kinetics}.
These observations suggest that the effects of activity cannot, in general, be characterized solely by propulsion strength, but depend crucially on the interplay between active and environmental timescales. In particular, environmental memory can qualitatively reshape the persistence of active motion. Since persistence plays a central role in many collective phenomena in active matter, memory effects may strongly influence the emergence of large-scale organization.

In this work we focus on a paradigmatic collective phenomenon of active matter, motility-induced phase separation (MIPS), in which purely repulsive self-propelled particles spontaneously separate into dense and dilute phases \cite{tailleur2008statistical,fily2012athermal,redner2013structure,cates2015motility,gonnella2015motility,caprini2020spontaneous}. 
MIPS emerges from the interplay between persistent propulsion and steric collisions: particles slow down in crowded regions, which promotes accumulation and self-trapping \cite{solon2018generalized, Sanoria.Nandi.2021}.
Since this mechanism is fundamentally controlled by persistence and collisional relaxation, it is expected to be highly sensitive to environmental memory. Recent studies have shown that viscoelastic and long-lived environmental correlations can strongly alter collective active dynamics, including enhanced clustering of Janus colloids in viscoelastic media \cite{dias2023environmental}, increased collective correlations in bacterial suspensions \cite{liao2023viscoelasticity, Liu.Wu.2021}, and the suppression of MIPS by hydrodynamic interactions \cite{matas2014hydrodynamic,zhou2026hydrodynamic} or inertia \cite{suma2014motility,mandal2019motility,mayo2026cooling}.
More generally, delayed interactions have recently been shown to strongly affect collective behavior in active systems with orientational alignment \cite{holubec2021finite,horton2025order}, further emphasizing the importance of memory and delay effects in nonequilibrium active matter. How environmental memory influences the onset and stability of MIPS, however, remains largely unexplored.

\begin{figure*}[t!]
    \centering
    \includegraphics[width=\linewidth]{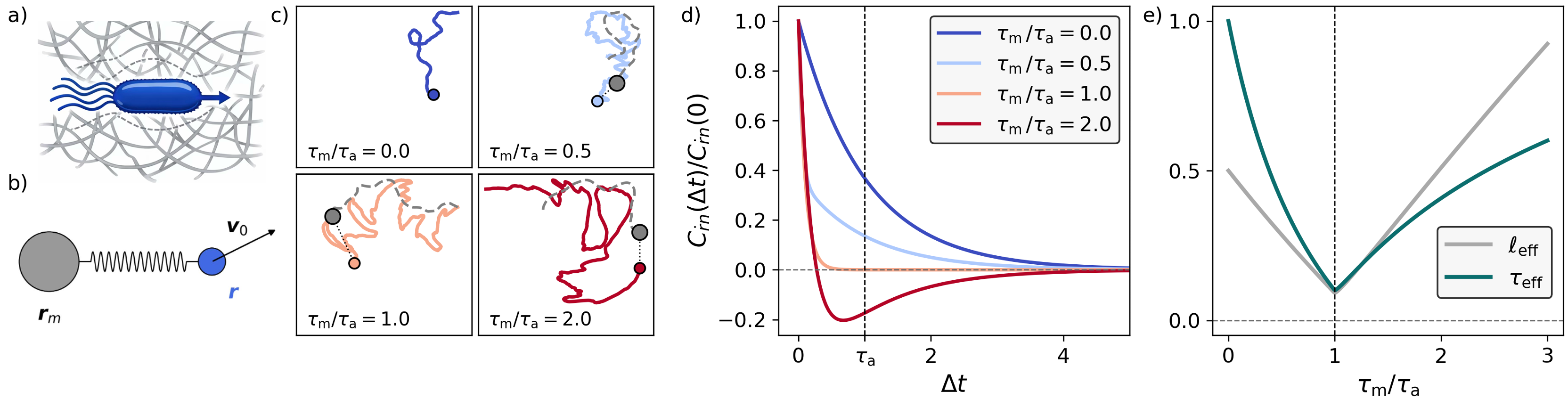}
    \caption{
Environmental memory reshapes active persistence through the competition between the active timescale $\tau_\mathrm{a}$ and the memory time $\tau_\mathrm{m}$. 
a) Schematic of an active particle moving in a viscoelastic environment.
b) Markovian embedding of the generalized Langevin dynamics: the active Brownian particle at position $\pmb{r}$ is coupled by a spring to an auxiliary memory coordinate $\pmb{r}_\mathrm{m}$ representing slowly relaxing environmental modes. 
c) Representative single-particle trajectories for different memory times $\tau_\mathrm{m}/\tau_\mathrm{a}$, showing both the physical particle and the auxiliary memory coordinate (gey dashed line). The case $\tau_\mathrm{m}/\tau_\mathrm{a}=0$ corresponds to the Markovian limit without memory. 
d) Velocity-orientation correlation function $C_{\dot r n}$ for different memory times. $\Delta t$ is in unit of $\tau_\mathrm{a}$. For sufficiently large $\tau_\mathrm{m}/\tau_\mathrm{a}$, delayed viscoelastic response generates negative correlations at intermediate times. 
e) Effective persistence time $\tau_\mathrm{eff}$ and effective persistence length $\ell_\mathrm{eff}$ as functions of $\tau_\mathrm{m}/\tau_\mathrm{a}$. Both quantities display a pronounced minimum around $\tau_\mathrm{m}=\tau_\mathrm{a}$, reflecting the competition between active persistence and delayed environmental response. 
$\tau_{\mathrm{eff}}$ is normalized by $\tau_\mathrm{a}$ and $\ell_{\mathrm{eff}}$ is normalized by $2v_0\tau_\mathrm{a}$ for optimized visualization. The minimum of $\tau_{\rm eff}$ decreases with the non-Markovian friction $\gamma_1$ (Fig.~\ref{fig:tau_eff_min}).}
    \label{fig:Fig1}
\end{figure*}

To address this question, we consider a coarse-grained description in which self-propelled particles with persistence time $\tau_{\rm a}$ move in a non-Markovian environment characterized by an additional memory timescale $\tau_{\rm m}$ through a generalized friction kernel. Within this framework, we investigate how environmental memory modifies the persistence of active motion, without relying on the microscopic details of a specific propulsion mechanism. The approach is relevant to systems in which the active drive remains approximately independent of the medium, such as synthetic Janus colloids and, more approximately, biological swimmers or motor-driven cargos in viscoelastic environments (Fig.~\ref{fig:Fig1}a). We show that the competition between active persistence and environmental memory produces a non-monotonic dependence of the effective persistence on $\tau_{\rm m}$, which in turn strongly affects collective behavior. As a consequence, motility-induced phase separation is suppressed when the active and environmental timescales become comparable and is recovered in the long-memory regime.

\section{Active particles with environmental memory}
\label{sec:model_intro}

To describe active motion in a viscoelastic environment, we employ a generalized Langevin equation (GLE) with delayed friction and thermal noise correlations~\cite{zwa61,mori1965transport,kubo1966fluctuation}. Such effective non-Markovian descriptions naturally emerge after integrating out slowly relaxing environmental degrees of freedom 
\cite{BackflowGLE,cui2018generalized,loos2020irreversibility,ayaz2022generalized,pelargonio2023generalized,busiello2024unraveling,hery2024derivation} and are widely used to model dynamics of colloids in viscoelastic baths \cite{metzler2000random,ViscGLE_Goy,di2020thermodynamic,DiTerGLE,doerries2021correlation,gomez2021work,ginot2022barrier,ginot2022recoil}. Particles are driven by an active propulsion velocity $\pmb v_{\rm a}^i(t)$ and interact via a repulsive force. Their dynamics obey the  GLE
\begin{equation}\label{eq:GLE_general}
\int_{-\infty}^t ds \,\gamma(t-s)\,\dot{\pmb r}^i(s)
=
\pmb F_{\mathrm{int}}^i(t)
+
\hat{\gamma}\,\pmb v_{\rm a}^i(t)
+
\boldsymbol{\eta}^i(t)\,.
\end{equation}
where $\gamma(t)$ is a memory kernel describing the delayed mechanical response of the environment, $\pmb F_{\rm int}^i$ is the repulsive interaction force acting on particle $i$, and $\boldsymbol{\eta}^i(t)$ is a Gaussian fluctuating force satisfying the fluctuation-dissipation relation
\begin{equation}
\langle \eta^i_\alpha(t)\eta^j_\beta(s)\rangle
=
\delta^{ij}\delta_{\alpha\beta}
k_{\rm B}T\,\gamma(|t-s|).
\end{equation}
The active velocities are assumed independent and isotropic, with correlations
\begin{equation}
\langle v_a^{i,\alpha}(t)v_a^{j,\beta}(s)\rangle
=
\delta^{ij}\delta^{\alpha\beta}
v_0^2\rho(|t-s|),
\end{equation}
where $\{\alpha,\beta\}$ indicate cartesian coordinates and $\rho(t)$ characterizes the temporal persistence of the active drive. Related models driven by colored noise include active Ornstein-Uhlenbeck particle \cite{szamel2014self,sevilla2019generalized}.

The active crawling velocity prefactor 
$\hat{\gamma}=\int_0^\infty dt\,\gamma(t)$
is chosen such that the long-time active transport is determined only by the activity itself. Indeed, in the absence of interactions, the mean-squared displacement obeys
\begin{equation}
\label{eq:msd_main}
\mathcal V_r^i(t)
=
\langle (r_t^i-r_0^i)^2\rangle
\overset{t\to\infty}{\simeq}
2(D^\eta+D^{\rm a})t,
\end{equation}
with
\begin{equation}
\label{eq:msd_main_2}
D^\eta=\frac{k_{\rm B}T}{\hat{\gamma}},
\qquad
D^{\rm a}=\frac{v_0^2\hat{\rho}}{2},
\end{equation}
where
\(
\hat{\rho}=\int_0^\infty dt\,\rho(t)
\)
is the integrated persistence of the active process (see Sec.~\ref{sec:msd_calculation}). Importantly, the active contribution $D^{\rm a}$ depends only on the active drive and not on the viscoelastic memory kernel. A similar result where transport properties are independent of the memory kernel holds for the simpler case of constant propulsion, $\pmb v_{\rm a}^i(t)=\pmb v_0$, for which the steady-state velocity is $\langle \dot{\pmb r} \rangle = \pmb v_0$. Hence, scaling the active velocity with $\hat{\gamma}$ isolates memory-induced changes in temporal correlations from trivial changes in long-time propulsion strength. Similar generalized active Brownian dynamics with memory were recently considered in Ref.~\cite{sprenger2022active}, while constant mean propulsion velocity across wide environmental variation has been observed in swimming bacterial colonies~\cite{martinez2014flagellated, liao2023viscoelasticity}. 

To model a viscoelastic environment with a single relaxation timescale, we consider the exponential memory kernel
\begin{equation}\label{eq:exp_mem_ker}
\gamma(t)
=
\gamma_0\delta(t)
+
\frac{\gamma_1}{\tau_{\rm m}}
e^{-t/\tau_{\rm m}}
\Theta(t),
\end{equation}
where $\tau_{\rm m}$ characterizes the environmental memory time and $\Theta$ is the Heaviside funtion. With this memory kernel, the long-time friction $\hat{\gamma}$ takes the form
\begin{equation}\label{eq:gamma_hat_def}
    \hat{\gamma} = \gamma_0 + \gamma_1.
\end{equation}
Eq.~\ref{eq:exp_mem_ker} continuously interpolates between three regimes: a Markovian fluid with friction coefficient $\hat{\gamma}$ in the limit $\tau_{\rm m} \to 0$, a viscoelastic Maxwell fluid for finite $\tau_{\rm m}$~\cite{ViscGLE_Goy, Paul.Banerjee.2018}, and a Markovian fluid with friction coefficient $\gamma_0$ in the limit $\tau_{\rm m} \to \infty$. 

The corresponding GLE admits an equivalent description via a Markovian embedding obtained by coupling each active particle to an auxiliary hidden degree of freedom representing slowly relaxing environmental modes (see Sec.~\ref{sec:Mark_mem}). A schematic illustration is shown in Fig.~\ref{fig:Fig1}b, where the physical particle at position $\pmb r$ is linearly coupled to an auxiliary coordinate $\pmb r_{\rm m}$, storing information about its past motion over the timescale $\tau_{\rm m}$. This representation provides both a transparent physical interpretation of the delayed response and an efficient framework for numerical simulations. For $\gamma_1 = \gamma_0$ and no interparticle interactions, the resulting dynamics is equivalent to that of elastic dumbbell particles commonly used to derive constitutive relations for viscoelastic polymeric fluids~\cite{bird1987dynamics}. Related active dumbbell models have similarly shown that competing relaxation timescales can strongly influence phase separation and collective dynamics~\cite{carenza2025arrested}.

\section{Single particle dynamics}

To identify the physical mechanism through which environmental memory reshapes collective active dynamics, we first analyze the motion of a single non-interacting active particle  ($\pmb{F}_{\rm int}^i=\pmb{0}$), and we specialize to the two-dimensional active Brownian dynamics 
\begin{equation}\label{eq:v_exp}  
\pmb v_{\rm a}^i(t)
=
v_0\pmb n(t),
\qquad
\pmb n(t)
=
(\cos\theta(t),\sin\theta(t)),
\end{equation}
with orientational dynamics
\begin{equation}
\dot\theta(t)
=
\sqrt{2D_\theta}\,\xi_\theta(t),
\qquad
\langle
\xi_\theta(t)\xi_\theta(t')
\rangle
=
\delta(t-t')\,
\end{equation}
and exponentially correlated propulsion direction,
\begin{equation}
\langle
\pmb n(t)\cdot\pmb n(0)
\rangle
=
e^{-|t|/\tau_{\rm a}},
\qquad
\tau_{\rm a}=D_\theta^{-1}.
\end{equation}

For $\tau_{\rm m}\gtrsim\tau_{\rm a}$, the particle undergoes persistent displacements followed by a reversal toward its previous position, as if pulled back by the delayed environmental response (red trajectories in Fig.~\ref{fig:Fig1}c). This behavior contrasts with the standard persistent motion of active Brownian particles (blue trajectories) and suggests that viscoelastic memory can generate anti-persistent dynamics opposing the active drive.

To characterize this effect quantitatively, we consider the velocity-orientation correlation function
\begin{equation}
C_{\dot r n}(t)
=
\left\langle
\dot{\pmb r}(t)\cdot \pmb n(0)
\right\rangle,
\label{eq:Crdotn_main}
\end{equation}
which measures how long the particle velocity remains aligned with its propulsion direction. For the exponential memory kernel of Eq.~\ref{eq:exp_mem_ker}, this correlation function can be calculated analytically (see Sec.~\ref{sec:2D_active_part}) and takes the form
\begin{equation}
\label{eq:Cvn_main}
C_{\dot r n}(t)
= v_0\left(
A e^{-t/\tau_{\rm a}} 
+
B e^{-t/\tau_{\rm v}}\right)
\end{equation}
where
\begin{equation}\label{eq:tau_v}
\tau_{\rm v}
=
\frac{\gamma_0\tau_{\rm m}}{\hat{\gamma}}
\end{equation}
is the viscoelastic relaxation timescale associated with the delayed environmental response, and $A$ and $B$ are dimensionless coefficients defined in Eq.~\ref{eq:An_Bn_def}. The dynamics is therefore controlled by the competition between the active persistence time $\tau_{\rm a}$ and the memory timescale $\tau_{\rm m}$, with a new viscoelastic relaxation time $\tau_{\rm v}$ emerging. The velocity-orientation correlation (Fig.~\ref{fig:Fig1}d) confirms the behavior observed in the trajectories: negative velocity-orientation correlations emerge near $t \sim \tau_{\rm a}$ for $\tau_{\rm m}>\tau_{\rm a}$, signaling transient motion opposite to the initial propulsion direction, and monotonic decay otherwise, consistent with standard active Brownian motion.

To quantify the net persistence of the trajectories despite the negative correlations, we define an effective persistence time
\begin{equation}\label{eq:tau_eff}
\tau_{\rm eff}
=
\int_0^\infty dt\,
\left|
\frac{C_{\dot r n}(t)}
{C_{\dot r n}(0)}
\right|,
\end{equation}
which corresponds to the active timescale $\tau_{\rm a}$ when $\tau_{\rm m}=0$. We also define the effective run length
\begin{equation}
\ell_{\rm eff}
=
\int_0^\infty dt\,
|C_{\dot r n}(t)|
=
v_{\rm eff}\tau_{\rm eff},
\label{eq:leff_veff}
\end{equation}
which in turn defines the effective velocity $v_{\rm eff} = \ell_{\rm eff} / \tau_{\rm eff}$, which takes the form
\begin{equation}\label{eq:v_eff_dilute}
   v_{\rm eff}=C_{\dot r n}(0) =
v_0\left(
1+
\frac{\gamma_1\tau_\mathrm{m}}
{\gamma_0(\tau_{\rm a}+\tau_\mathrm{m})}
\right)\,,
\end{equation}
with $v_{\rm eff}=v_0$ in the Markovian case. This effective velocity is instantaneous and larger in the presence of memory, but does not affect the long time diffusion as prescribed in Eq.~\ref{eq:msd_main}; particles move more quickly at short bursts, but viscoelastic effects dampen and constrain diffusion on longer timescales. 

As shown in Fig.~\ref{fig:Fig1}e, both the effective persistence time and the effective run length display a pronounced minimum when $\tau_{\rm m}\sim\tau_{\rm a}$, where delayed stresses most strongly oppose orientational persistence. This non-monotonic dependence anticipates a corresponding re-entrant collective response.

\section{Memory-controlled phase separation}

\begin{figure*}
    \centering
    \includegraphics[width=0.99\linewidth]{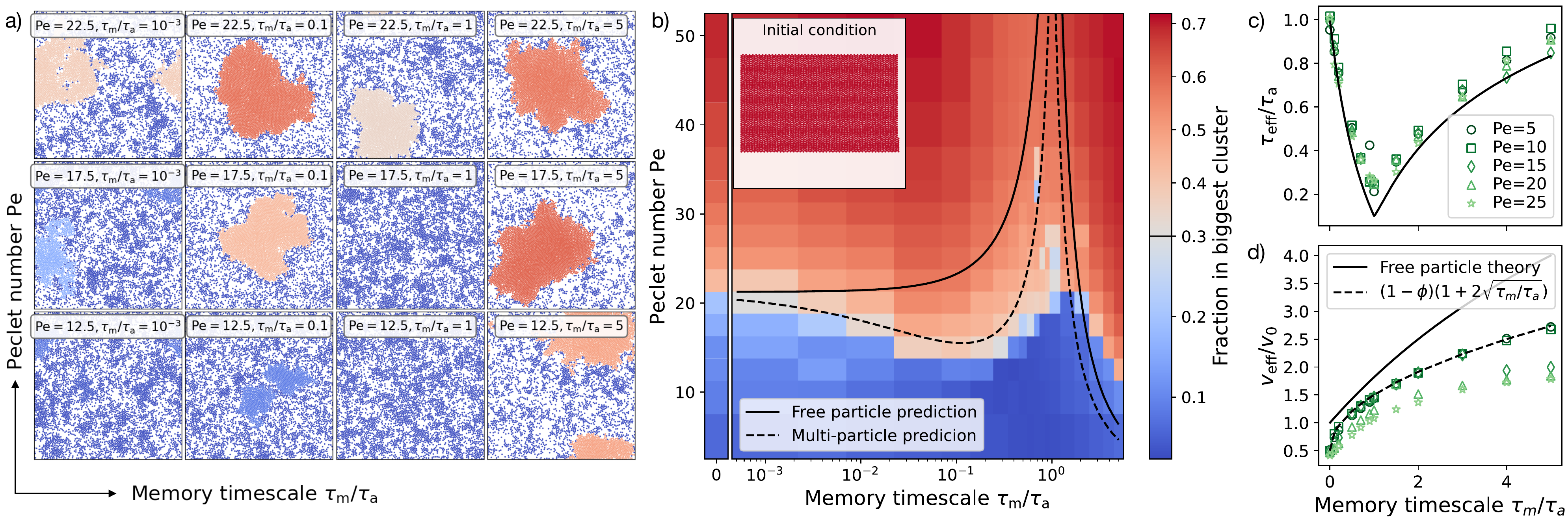}
    \caption{Collective dynamics of interacting active particles in a viscoelastic medium.  a) Representative steady states for different memory times $\tau_\mathrm{m}/\tau_\mathrm{a}$ and Péclet numbers $\mathrm{Pe}$ for $5000$ particles at packing fraction $0.5$, $\gamma_0=0.1$ and $\gamma_1=0.9$. Clustering is suppressed around $\tau_\mathrm{m}\sim\tau_\mathrm{a}$ and reappears at larger memory times.  b) Fraction of particles in the system's largest cluster as a function of $\tau_\mathrm{m}/\tau_\mathrm{a}$ and $\mathrm{Pe}$. Black solid and dashed lines show the theoretical prediction for the transition from the MIPS to gas phase by the contour of the effective Péclet number 
    (Eq.~\ref{eq:peclet_eff}) with $v_{\rm eff}$ given by Eq.~\ref{eq:leff_veff} (single particle, solid line) or Eq.~\ref{eq:veff_multiparticle} (multi-particle, dashed line). $\tau_{\rm eff}$ is given in Eq.~\ref{eq:leff_veff} for both cases. (inset) At t=0, all particles belong to a dense rectangular cluster. c) Effective persistence time $\tau_{\rm eff}$ measured in the interacting system for different $\mathrm{Pe}$, compared with the free-particle prediction (solid line).  d) Effective velocity $v_{\rm eff}$ of single particles (solid line) and in interacting systems $\bar{v}_{\rm eff}$ as a function of memory time. Symbols correspond to simulations and the dashed line shows the predicted square-root scaling induced by interactions.
}
    \label{Fig2}
\end{figure*}

To investigate how viscoelastic memory reshapes motility-induced phase separation, we consider interacting active particles with repulsive soft interactions
\begin{equation}\label{eq:interaction}
\pmb F_{\rm int}^i
=
\kappa
\sum_{j\neq i}
(\sigma -|\pmb r_{ij}|)\,
\Theta(\sigma-|\pmb r_{ij}|)\hat{\pmb r}_{ij}
\end{equation}
where $\sigma$ is the particle diameter, and $\hat{\pmb r}_{ij}$ is a unit vector pointing along $\pmb r_{ij} = \pmb r_i - \pmb r_j$. $\kappa$ is the stiffness of the repulsive potential, and the Heaviside function $\Theta$ enforces purely repulsive interactions. 
Steady-state properties are obtained numerically using the Markovian embedding introduced above (see Sec~\ref{sec:supp_simulation_details} for simulation details). In the Markovian limit, activity is characterized by the Péclet number
\begin{equation}
{\rm Pe}
=
\frac{v_0\tau_{\rm a}}{\sigma},
\end{equation}
which compares the persistence length of the active motion to the particle size.

As shown in Fig.~\ref{Fig2}a, viscoelastic memory produces a re-entrant collective behavior in which MIPS is suppressed around $\tau_{\rm m}\sim\tau_{\rm a}$ and restored at larger memory times. Starting from an initially phase-separated configuration, the system behaves similarly to standard active Brownian dynamics when $\tau_{\rm m}\approx 0$, while clustering destabilizes when the viscoelastic relaxation becomes comparable to the active persistence time. At larger memory times, however, phase separation reappears and emerges for Péclet numbers below the Markovian transition threshold (for instance $\mathrm{Pe}=12.5$ at $\tau_{\rm m}/\tau_{\rm a}=5$). The full phase diagram of the cluster fraction, shown in Fig.~\ref{Fig2}b, confirms the existence of an intermediate anti-persistent regime separating two phase-separated regions.
When $\tau_\mathrm{m}/\tau_\mathrm{a}\gg1$, the dynamics approaches an effective Markovian limit controlled by the instantaneous friction coefficient $\gamma_0$, leading to the saturation of the phase boundary at finite Péclet number (see Supplementary Fig.~\ref{fig:large_deborah}).

Since the onset of MIPS is controlled by persistence and run length, based on the single-particle results of Sec.~\ref{sec:model_intro}, we posit that the observed phenomenology could be captured by the effective Péclet number
\begin{equation}
{\rm Pe}_{\rm eff}
=
\frac{v_{\rm eff}\tau_{\rm eff}}{\sigma},
\label{eq:peclet_eff}
\end{equation}
where $v_{\rm eff}$ and $\tau_{\rm eff}$ are obtained from the single particle velocity-orientation correlation (see Eqs.~\ref{eq:tau_eff},\ref{eq:leff_veff}). In Fig.~\ref{Fig2}b, we plot as contour lines ${\rm Pe}_{\rm eff}={\rm Pe}^*$ (solid line) for the theoretical prediction of the transition, where ${\rm Pe}^*$ denotes the critical Péclet number for phase separation in the Markovian limit (see Figure \ref{fig:sup_cluster_threshold}). This simple criterion captures the suppression of MIPS around $\tau_{\rm m}\sim\tau_{\rm a}$, induced by the reduction of the effective persistence time $\tau_{\rm eff}$. Small but systematic deviations nevertheless remain at short memory times, indicating that interactions modify the effective dynamical scales controlling phase separation. Figure~\ref{Fig2}c shows, however, that the effective persistence time in the interacting system measured from simulations remains remarkably close to the free-particle prediction across the full parameter range explored. The remaining discrepancies must therefore originate from interaction-induced renormalization of the effective propulsion velocity $v_{\rm eff}$.

We therefore measure the effective velocity
\begin{equation}
    \bar{v}_{\rm eff} = C^{\rm int}_{\dot{r}n}(0)
\end{equation}
derived from the velocity-orientation correlation function in the interacting system $C^{\rm int}_{\dot{r}n}$. We focus on particles that remain in the homogeneous phase ($\mathrm{Pe}<\mathrm{Pe}^*$), corresponding to the darker symbols in Fig.~\ref{Fig2}d. In the Markovian limit, interactions reduce the effective velocity $\bar{v}_{\rm eff}(\tau_{\rm m}=0)\approx (1-\phi)v_0$, where $\phi$ is the packing fraction. For small memory times, the increase of $\bar{v}_{\rm eff}$ relative to this Markovian value follows approximately a square-root dependence on $\tau_{\rm m}$, in contrast to the free-particle prediction of Eq.~\eqref{eq:v_eff_dilute}, which scales linearly with $\tau_{\rm m}$.
This faster increase of the effective velocity relative to the free particle case transiently lowers the MIPS transition line. We incorporate the measured $\bar{v}_{\rm eff}$ (dashed line) of Fig.~\ref{Fig2}d into ${\rm Pe}_{\rm eff}$, and determine the corresponding transition line in the phase space by matching to the Markovian case. Note that this procedure takes into account the $1-\phi$ rescaling which is also present in the Markovian case. This shifts the phase boundary toward lower Péclet numbers at small $\tau_{\rm m}$, yielding an improved agreement with the simulations (dashed line in Fig.~\ref{Fig2}b).

To understand the effect of interactions on the effective velocity, we note that $\bar{v}_{\rm eff}$ can be written as
\begin{equation}\label{eq:C_int_def}
\bar{v}_{\rm eff}
= C_{\dot{r}n}^{\rm \, int}(0) = 
C_{\dot{r}n}^{\rm \, free}(0)
+
\int_0^\infty ds\,
\mu(s)\,
C_{nF}(s),
\end{equation}
where we have dropped the particle index for simplicity, $\mu(t)$ is the mobility kernel defined as 
\begin{equation}
    \mu(t)=\frac{1}{\gamma_0}\delta(t)-\frac{\gamma_1}{\gamma_0^2\tau_\mathrm{m}}e^{-t/\tau_{\rm v}}\Theta(t),
\end{equation}
and
\begin{equation}
C_{nF}(s)
=
\left\langle
\pmb n(s)\cdot \pmb F_{\rm int}(0)
\right\rangle
\end{equation}
measures the correlation between the propulsion direction and the interaction forces. For repulsive interactions, $C_{nF}(0)<0$, as collisions typically oppose self-propulsion. Moreover, the orientation-force correlation decays over the orientational persistence time,
\begin{equation}
C_{nF}(s)
=
C_{nF}(0)e^{-s/\tau_{\rm a}},
\end{equation}
so that interactions inherit the same temporal structure as the active dynamics itself (see \ref{sec:CnF_calculations}). From Eq. \ref{eq:C_int_def} one obtains
\begin{equation}
C_{\dot{r}n}^{\rm \, int}(0)
=
C_{\dot{r}n}^{\rm \, free}(0)
+
\frac{C_{nF}(0)}{\gamma_0}
\left(
1-
\frac{\gamma_1 \tau_{\rm a}}
{\hat{\gamma}(\tau_{\rm a}+\tau_{\rm v})}
\right),
\end{equation}
with $\tau_{\rm v}$ defined in Eq.~\ref{eq:tau_v}. As $C_{nF}(0)<0$ in all analyzed cases, interactions generate a negative correction to the velocity-orientation correlation, and thus $\bar{v}_{\rm eff}$.

To further characterize this correction, we measure the equal-time orientation-force correlation $C_{nF}(0)$ for different parameter values and in the homogeneous regime where clustering remains weak. We find that its amplitude develops a strong dependence on both memory and density, and is well approximated by
\begin{equation}
C_{nF}(0)
\simeq
\phi
\left(
-\hat{\gamma}
+
\frac{
b\sqrt{\tau_\mathrm{m}}
}{
1+\tau_\mathrm{m}/\tau_{nF}
}
\right),
\end{equation}
where $\phi$ is again the packing fraction, $b$ and $\tau_{nF}$ are fitting parameters where $\tau_{nF}$ characterizes the decorrelation time of the interaction forces (see Fig.~\ref{fig:CnF0_fits}). The prefactor $\phi$ reflects the increasing role of collisions at larger densities. 
Substituting this empirical form into the previous expression yields, for small viscoelastic relaxation times $\tau_{\rm m}/\tau_\mathrm{a}\ll1$,
\begin{equation}\label{eq:veff_multiparticle}
\bar{v}_{\rm eff}
\approx
v_0(1-\phi)
\left(
1+
b\sqrt{\tau_\mathrm{m}/\tau_\mathrm{a}}
\right),
\end{equation}
with $b\approx2$ for $\phi=0.5$, as shown in Fig.~\ref{Fig2}d. After rescaling by $1-\phi$, the effective velocity therefore grows more rapidly at small memory times than predicted by the dilute theory Eq.~\ref{eq:v_eff_dilute}. The square-root behavior reflects the collective nature of the interaction-induced renormalization of the propulsion response, since even weak viscoelastic delays modify repeated collisional encounters and transient caging dynamics. 

\section{Metastability and slow nucleation}\label{sec:metastability}

\begin{figure}[t!]
    \centering
    \includegraphics[width=\linewidth]{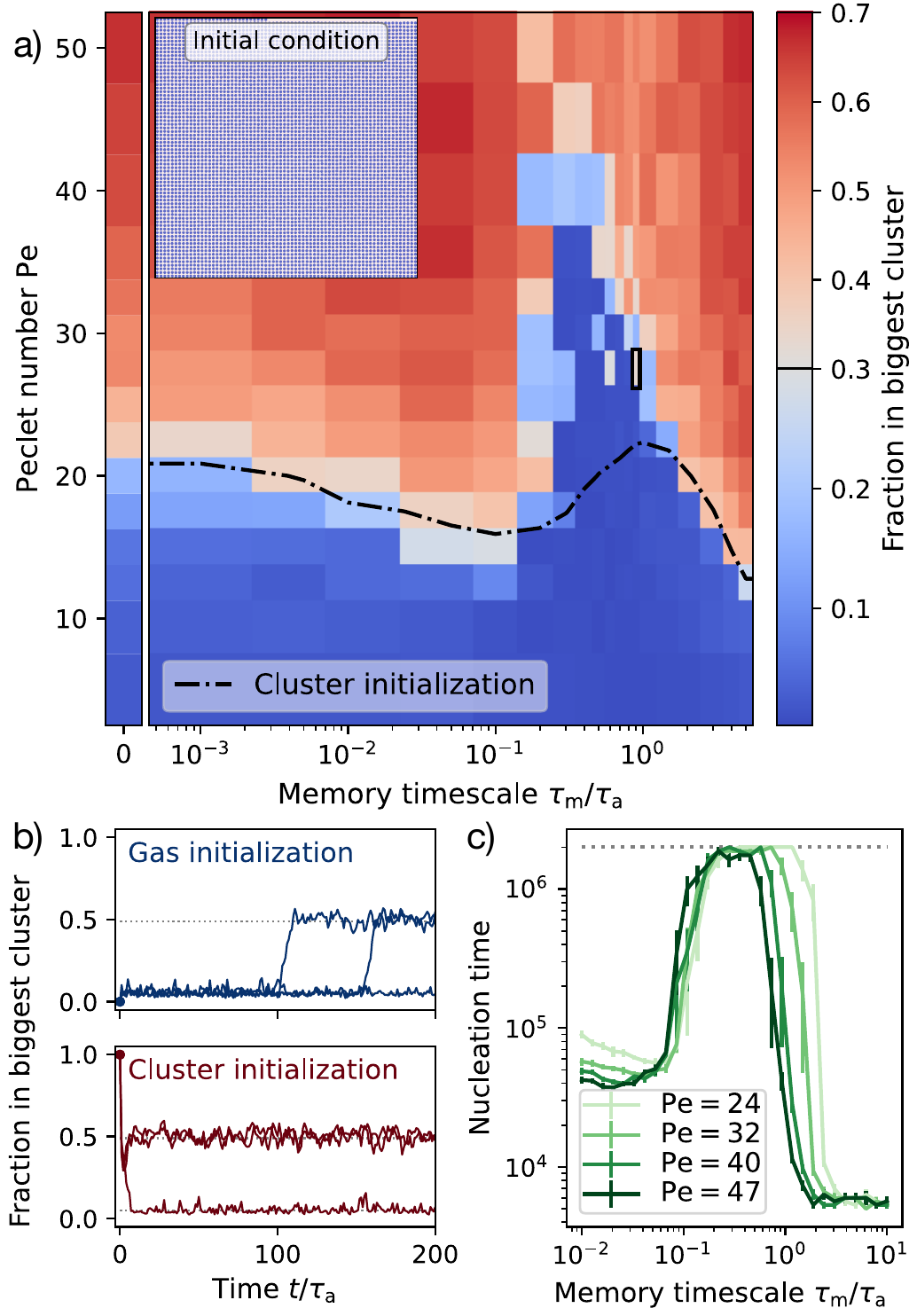}
    \caption{
Dynamics and metastability of viscoelastic MIPS. 
a) Phase diagram obtained from homogeneous gas-like initial conditions, and identical parameters as in Fig.~\ref{Fig2}b. Here, the suppression region is broader and shifted toward smaller $\tau_\mathrm{m}$, although it remains centered around $\tau_\mathrm{m}\sim\tau_\mathrm{a}$, indicating the importance of kinetic nucleation effects in the formation of the dense phase. (\emph{inset}) The initial state is fully dilute.
b) Time evolution of the cluster fraction for different stochastic realizations in the metastable regime at identical parameter values ($\mathrm{Pe}=27.5$ and $\tau_\mathrm{m}/\tau_\mathrm{a}=0.9$), marked with a rectangle in panel a. Depending on the trajectory, the system may remain phase separated, dissolve into a homogeneous state, undergo phase separation, or remain dilute throughout the dynamics.
c) Mean nucleation time (logscale, in units of $\tau_\mathrm{a}$) for MIPS as a function of memory time and activity. The nucleation time displays a pronounced maximum at $\tau_\mathrm{m}\sim\tau_\mathrm{a}$, signaling a strong slowing down of the collective dynamics near the metastable transition region. Nucleation time is capped at the simulation duration, indicated by the grey dotted line. 
}
    \label{fig:dynamics}
\end{figure}

To further characterize the dynamical regime identified in Fig.~\ref{Fig2}, we now investigate the formation of clusters starting from homogeneous gas-like initial conditions. The resulting phase diagram is shown in Fig.~\ref{fig:dynamics}a. Compared to the stability diagram obtained from initially phase-separated states, the region where MIPS is suppressed becomes broader and shifted toward slightly smaller memory times, although the transition still occurs for $\tau_\mathrm{m}$ of the order of $\tau_\mathrm{a}$. 
This difference indicates that the emergence of phase separation is controlled not only by the stability of the dense phase, but also by the kinetics of nucleation and cluster growth \cite{stenhammar2013continuum}.
In particular, the sensitivity to the preparation protocol already suggests the presence of slow relaxation dynamics and competing timescales near $\tau_\mathrm{m}\sim\tau_\mathrm{a}$. 

This metastability is illustrated directly in Fig.~\ref{fig:dynamics}b, which shows the time evolution of the cluster fraction for several independent stochastic realizations within the transition region. Depending on fluctuations, the system may remain homogeneous for the entire simulation, nucleate a dense cluster that subsequently dissolves, or evolve toward a long-lived phase-separated state. The coexistence of these qualitatively different dynamical trajectories confirms that the intermediate-memory regime is characterized by competing timescales. Additional trajectories across the phase diagram, shown in Fig.~\ref{fig:supp_dynamics}a, exhibit the same phenomenology.

To quantify this dynamical slowdown, we measure in Fig.~\ref{fig:dynamics}c the characteristic time required for the system to reach a phase-separated state. The nucleation time develops a pronounced maximum around $\tau_\mathrm{m}\sim\tau_\mathrm{a}$, precisely where the effective persistence time is minimized. In this regime, delayed viscoelastic stresses strongly reduce the persistence of directed motion, weakening the self-trapping mechanism responsible for MIPS. As a consequence, collisions are no longer able to sustain stable dense nuclei over persistence times, and cluster formation becomes dramatically slower. The peak in the nucleation time therefore reflects the competition between active persistence and delayed viscoelastic relaxation. The full phase diagram of transition times, shown in Supplementary Fig.~\ref{fig:supp_dynamics}b, further confirms that relaxation becomes significantly slower near the phase boundary. In Fig.~\ref{fig:suppfig_fric_ratios}, we show that the kinetic suppression of MIPS also occurs for different values of $\gamma_0$ and $\gamma_1$, indicating that this transition persists across varying strengths of viscoelastic memory. 

Taken together, these results show that environmental memory reshapes MIPS through the competition between memory-induced anti-persistent motion and enhanced propulsion at long memory times. Around $\tau_{\rm m}\sim\tau_{\rm a}$, this competition generates a broad metastable regime with competing dynamical pathways and larger nucleation times, highlighting that memory modifies not only the stability of phase-separated states but also their formation kinetics.

\section*{Discussion}

In this work, we studied active particles embedded in viscoelastic environments using a generalized Langevin description with memory. We showed that the competition between the active persistence time $\tau_\mathrm{a}$ and the viscoelastic relaxation time $\tau_\mathrm{m}$ reshapes the persistence of active motion. Delayed viscoelastic stresses generate anti-persistent dynamics and suppress the effective persistence when $\tau_\mathrm{m}\sim\tau_\mathrm{a}$, leading to strong inhibition of motility-induced phase separation and the emergence of a broad metastable regime with slow nucleation dynamics. At longer memory times, the reduced short-time friction enhances propulsion and restores clustering. Since viscoelastic relaxation is common in both biological and synthetic systems, these findings suggest that environmental memory may play an important role in determining collective behavior across a broad range of systems. More generally, our results highlight how the interplay between active and environmental timescales can shape collective organization far from equilibrium.

We observe here that the non-monotonic dependence of phase separation on the viscoelastic memory time might be expected from the behavior of the medium in which our particles are effectively embedded. The exponential memory kernel considered in Eq.~\ref{eq:exp_mem_ker} reproduces that of a Maxwell fluid~\cite{ViscGLE_Goy, Paul.Banerjee.2018}. Maxwell fluids are characterized by an oscillatory loss modulus $G''(\omega)$ at driving frequency $\omega$ peaks when $\omega = 1/\tau^*$~\cite{bird1987dynamics}, where $\tau^*$ is the characteristic relaxation time in the material. Identifying $\tau_{\rm m}$ in our model with $\tau^*$ and $\tau_{\rm a}$ with $\omega^{-1}$ suggests that the suppression of MIPS occurs when the material has the highest effective loss modulus $G''$. The medium might therefore be dissipating energy most strongly in this regime, and future studies analyzing how medium rheology, energetics and response affect active persistence  may improve our understanding of the effect of complex media on active matter. 

Finally, an important open direction concerns more realistic forms of environmental memory~\cite{quevedo2025active,karmakar2026beyond} and their collective consequences~\cite{puljiz2019memory}. In the present work the viscoelastic response is assumed spatially local and characterized by a single relaxation timescale, such that each particle couples independently to an exponentially relaxing environment. Real complex fluids, however, often exhibit broad spectra of relaxation times and long-lived correlations that are more naturally described by multi-timescale or power-law memory kernels \cite{molina2018crossover,metzler2000random,chechkin2009fluctuation,goychuk2014molecular,chechkin2017brownian,vitali2018langevin,khadem2022stochastic}. In addition, stresses generated by particle motion can propagate through the medium and persist over finite distances, producing correlated memory forces and retarded many-body interactions between active particles. A possible approach to incorporating spatial correlations would be to employ a field-theoretic description of the viscoelastic medium \cite{dean2011diffusion,demery2014generalized,venturelli2023memory,venturelli2024stochastic,pruszczyk2025recoil}. Extending active matter theories to such nonlocal and broadly distributed viscoelastic memory may reveal collective regimes beyond conventional MIPS phenomenology, including memory-mediated synchronization, delayed clustering, anomalous transport, or cooperative active flows.

\begin{acknowledgments}
\noindent The authors thank  Ylann Rouzaire, Vincenzo Maria Schimmenti and Matteo Ciarchi for useful discussions and comments on the manuscript. The authors also acknowledge discussions with Alexis Poncet, who was independently studying a closely related system at the time the authors prepared this manuscript.
\end{acknowledgments}

\section*{Authors contributions}
\noindent I.D.T., L.K. and J.D.T. designed, planned, and conducted the research;
L.K. and J.D.T. designed simulations and performed the numerical work, I.D.T. performed analytical calculations;
I.D.T., L.K. and J.D.T. wrote the manuscript.  L.K. was supported by the MSCA Postdoctoral fellowship. 

\section*{Code availability}

Code to run simulations: \href{https://github.com/jacktreado/MaxwellABP}{MaxwellABP}

\bibliography{biblio, jack_biblio}

\clearpage

\onecolumngrid

\setcounter{equation}{0}
\setcounter{figure}{0}
\setcounter{table}{0}
\setcounter{page}{1}
\setcounter{section}{0}
\makeatletter
\renewcommand{\theequation}{S\arabic{equation}}
\renewcommand{\thefigure}{S\arabic{figure}}
\setcounter{secnumdepth}{2} 
\renewcommand{\thesection}{S\arabic{section}} 

\begin{center}
{\LARGE Supplemental Material for}\\
\vspace{0.4cm}
\textbf{\Large Passive memory reshapes active persistence}\\
\hspace{0.5cm}
\end{center}

\begin{center}

{\large
Ivan Di Terlizzi$^{1,2,*}$, 
Lara Koehler$^{1,3,*}$, 
John D. Treado$^{1,3,*}$
}

\vspace{0.3cm}

{\small
$^{1}$Max Planck Institute for the Physics of Complex Systems, 
Nöthnitzer Straße 38, 01187 Dresden, Germany\\
$^{2}$Ludwig-Maximilians-Universit\"at M\"unchen, 
Arnold-Sommerfeld-Center for Theoretical Physics, 
M\"unchen, Germany\\
$^{3}$Cluster of Excellence Physics of Life, TU Dresden, 
01307 Dresden, Germany
}

\vspace{0.2cm}

{\small
$^{*}$These authors contributed equally to this work
}

\end{center}

\section{Model details}\label{sec:model_details}

In this section, we derive the analytical results used throughout the main text. We derive the long-time mean-square displacement Eqs.~\ref{eq:msd_main},\ref{eq:msd_main_2} (Sec.~\ref{sec:msd_calculation}), the velocity-orientation correlation function Eq.~\ref{eq:Cvn_main} for two-dimensional active Brownian particles (Sec.~\ref{sec:2D_active_part}), and the effective persistence time shown in Fig.~\ref{fig:Fig1}e (Sec.~\ref{sec:tau_eff_calculation}). We also present the Markovian embedding formalism associated with Fig.~\ref{fig:Fig1}b and used to perform the numerical simulations (Sec.~\ref{sec:Mark_mem}).

\subsection{Long-time scaling of the mean-square displacement}
\label{sec:msd_calculation}

Here we show that the prefactor $\hat{\gamma}$ multiplying the active velocity in the generalized Langevin equation ensures that the long-time active contribution to the mean-square displacement remains independent of the rheological properties of the environment. As a consequence, the asymptotic active diffusivity is controlled only by the statistics of the active drive and not by the details of the memory kernel.

\noindent
We consider the $d$-dimensional overdamped generalized Langevin equation
\begin{equation}
\int_{-\infty}^{t}\gamma(t-s)\,\dot{\pmb r}_s\,ds
=
\hat{\gamma}\,\pmb v_t+\boldsymbol{\eta}_t,
\label{eq:gle_vector_general}
\end{equation}
where $\pmb r_t\in\mathbb R^d$, $\gamma(t)$ is a causal scalar memory kernel, and
\begin{equation}
\hat{\gamma}\equiv \int_0^\infty \gamma(t)\,dt
\end{equation}
is its time integral. The thermal noise is assumed Gaussian, stationary, isotropic, and to satisfy the fluctuation-dissipation relation componentwise,
\begin{equation}
\langle \eta_t^\alpha \eta_0^\beta\rangle
=
\delta_{\alpha\beta}\,k_{\rm B}T\,\gamma(|t|),
\qquad \alpha,\beta=1,\dots,d.
\label{eq:FDT_vector}
\end{equation}
The active drive $\pmb v_t$ is taken to be a stationary isotropic process, independent of $\boldsymbol{\eta}_t$, with zero mean and correlations
\begin{equation}
\langle v_t^\alpha v_0^\beta\rangle
=
v_0^2\,\rho^{\alpha\beta}(t),
\qquad
\rho^{\alpha\beta}(t)=\delta_{\alpha\beta}\rho(t).
\label{eq:rho_isotropic}
\end{equation}
Equivalently,
\begin{equation}
\langle \pmb v_t\cdot \pmb v_0\rangle=d\,v_0^2\,\rho(t).
\end{equation}
Introducing the frequency-dependent mobility
\begin{equation}
\hat{\mu}(\omega)=\frac{1}{\hat{\gamma}(\omega)},
\qquad
\hat{\gamma}(\omega)\equiv \int_0^\infty \gamma(t)e^{-i\omega t}\,dt,
\end{equation}
Eq.~\ref{eq:gle_vector_general} becomes, componentwise,
\begin{equation}
\dot r^\alpha(\omega)
=
\hat{\mu}(\omega)\eta^\alpha(\omega)
+
\hat{\gamma}\,\hat{\mu}(\omega)\,v^\alpha(\omega).
\label{eq:vel_freq_general}
\end{equation}
The velocity correlation tensor therefore splits into passive and active parts,
\begin{equation}
\hat C_{\dot r}^{\alpha\beta}(\omega)
=
\hat C_{\dot r,\eta}^{\alpha\beta}(\omega)
+
\hat C_{\dot r,{\rm a}}^{\alpha\beta}(\omega).
\end{equation}
Using Eq.~\ref{eq:FDT_vector}, one obtains
\begin{equation}
\hat C_{\dot r,\eta}^{\alpha\beta}(\omega)
=
\delta_{\alpha\beta}\,2k_{\rm B}T\,\Re \hat{\mu}(\omega),
\label{eq:Cv_passive_tensor}
\end{equation}
while the active part reads
\begin{equation}
\hat C_{\dot r,{\rm a}}^{\alpha\beta}(\omega)
=
\hat{\gamma}^{\,2}\,|\hat{\mu}(\omega)|^2\,\hat C_v^{\alpha\beta}(\omega),
\label{eq:Cv_active_tensor_general}
\end{equation}
where
\begin{equation}
\hat C_v^{\alpha\beta}(\omega)
=
\int_{-\infty}^{\infty}dt\,e^{i\omega t}\langle v_t^\alpha v_0^\beta\rangle
=
\delta_{\alpha\beta}\,v_0^2\hat\rho(\omega).
\label{eq:Cvdrive_tensor}
\end{equation}
Hence, by isotropy,
\begin{equation}
\hat C_{\dot r}^{\alpha\beta}(\omega)
=
\delta_{\alpha\beta}\,\hat C_{\dot r}(\omega),
\end{equation}
with single-component spectrum
\begin{equation}
\hat C_{\dot r}(\omega)
=
2k_{\rm B}T\,\Re \hat{\mu}(\omega)
+
\hat{\gamma}^{\,2}|\hat{\mu}(\omega)|^2\,v_0^2\hat\rho(\omega).
\label{eq:Cv_component_general}
\end{equation}
It is convenient to characterize long-time diffusion through the Green-Kubo relation. For one Cartesian component,
\begin{equation}
D_{\rm eff}
=
\int_0^\infty dt\,\langle \dot r^\alpha_t \dot r^\alpha_0\rangle
=
\frac{1}{2}\hat C_{\dot r}(0),
\label{eq:GreenKubo_component}
\end{equation}
provided the integral converges. Since $\hat\mu(0)=1/\hat\gamma$, Eq.~\ref{eq:Cv_component_general} gives
\begin{equation}
D_{\rm eff}=D^\eta+D^{\rm a},
\end{equation}
with
\begin{equation}
D^\eta=\frac{k_{\rm B}T}{\hat\gamma},
\qquad
D^{\rm a}=\frac{v_0^2}{2}\hat\rho(0)
=
v_0^2\int_0^\infty \rho(t)\,dt.
\label{eq:Deff_general_isotropic}
\end{equation}
Thus, for each coordinate, the passive contribution retains the Einstein form, while the active contribution depends only on the integrated correlation of the propulsion process and is independent of the memory kernel. 

For the full displacement vector, isotropy implies
\begin{equation}
\langle |\pmb r(t)-\pmb r(0)|^2\rangle
\simeq 2d\,D_{\rm eff}\,t
\qquad (t\to\infty).
\label{eq:msd_vector_general}
\end{equation}

\subsection{Two-dimensional active Brownian particle}\label{sec:2D_active_part}

In this section, we study the two-dimensional active Brownian particle evolving in a viscoelastic environment with exponential memory kernel. We show that the delayed environmental response generates a second relaxation mode $\tau_{\rm v}$ and leading to the emergence of negative velocity-orientation correlations when $\tau_{\rm m}\sim\tau_{\rm a}$. We also derive the memory dependence of the effective velocity and persistence quantities used in the main text.

We specialize to the two-dimensional case in which the active velocity has fixed magnitude and diffusing orientation,
\begin{equation}
\pmb v_t=v_0\pmb n_t,
\qquad
\pmb n_t=(\cos\theta_t,\sin\theta_t),
\qquad
\dot\theta_t=\sqrt{2D_\theta}\,\xi_t,
\label{eq:abp_orientation}
\end{equation}
with $\xi_t$ a unit white noise. Writing
\begin{equation}
\tau_{\rm a}=D_\theta^{-1},
\end{equation}
the orientational process is stationary and isotropic, with
\begin{equation}
\langle n_t^\alpha n_0^\beta\rangle
=
\frac{\delta_{\alpha\beta}}{2}e^{-|t|/\tau_{\rm a}},
\qquad \alpha,\beta\in\{x,y\}.
\label{eq:ABP_iso_corr}
\end{equation}
Therefore,
\begin{equation}
\rho^{\alpha\beta}(t)=\delta_{\alpha\beta}\rho(t),
\qquad
\rho(t)=\frac{1}{2}e^{-|t|/\tau_{\rm a}},
\end{equation}
and
\begin{equation}
\hat\rho(\omega)=\frac{\tau_{\rm a}}{1+\omega^2\tau_{\rm a}^2}.
\label{eq:rhohat_abp}
\end{equation}
The single-component active-drive spectrum is thus
\begin{equation}
\hat C_v^{\alpha\beta}(\omega)
=
\delta_{\alpha\beta}\,
\frac{v_0^2\tau_{\rm a}}{1+\omega^2\tau_{\rm a}^2}.
\end{equation}
Substituting into Eq.~\ref{eq:Cv_component_general}, the velocity spectrum for each Cartesian component becomes
\begin{equation}
\hat C_{\dot r}(\omega)
=
2k_{\rm B}T\,\Re \hat\mu(\omega)
+
\hat\gamma^{\,2}|\hat\mu(\omega)|^2
\frac{v_0^2\tau_{\rm a}}{1+\omega^2\tau_{\rm a}^2}.
\label{eq:Cv_component_abp}
\end{equation}
The corresponding effective diffusion coefficient for one coordinate is
\begin{equation}
D_{\rm eff}
=
\int_0^\infty dt\,\langle \dot r^i_t\dot r^i_0\rangle
=
\frac{k_{\rm B}T}{\hat\gamma}
+\frac{v_0^2\tau_{\rm a}}{2},
\label{eq:Deff_abp_GK}
\end{equation}
and therefore
\begin{equation}
\langle |\pmb r(t)-\pmb r(0)|^2\rangle
\simeq
4D_{\rm eff}t,
\qquad
D_{\rm eff}
=
\frac{k_{\rm B}T}{\hat\gamma}
+\frac{v_0^2\tau_{\rm a}}{2}.
\label{eq:MSD_abp_final}
\end{equation}
As in the general discussion above, the active contribution
\begin{equation}
D^{\rm a}=\frac{v_0^2\tau_{\rm a}}{2}
\end{equation}
is independent of the memory kernel, while the latter controls the full time dependence of the velocity correlations.

A useful quantity to characterize the intermediate-time dynamics is the correlation between the particle velocity and the propulsion direction,
\begin{equation}
C_{\dot r n}(t)\equiv \langle \dot{\pmb r}_t\cdot \pmb n_0\rangle.
\label{eq:Crdotn_def_app}
\end{equation}
Using the causal mobility kernel $\mu(t)$, defined by
\begin{equation}
\dot{\pmb r}_t
=
\int_{-\infty}^{t}ds\,\mu(t-s)\Big[\hat{\gamma}v_0\,\pmb n_s+\boldsymbol{\eta}_s\Big],
\end{equation}
together with the independence of $\boldsymbol{\eta}$ and $\pmb n$, one finds
\begin{equation}\label{eq:C_rdotn_def}
C_{\dot r n}(t)
=
\hat{\gamma}v_0
\int_{-\infty}^{t}ds\,\mu(t-s)\,\langle \pmb n_s\cdot \pmb n_0\rangle .
\end{equation}
For the two-dimensional ABP dynamics,
\begin{equation}
\langle \pmb n_t\cdot \pmb n_0\rangle=e^{-|t|/\tau_{\rm a}},
\end{equation}
so that, after the change of variable $u=t-s$,
\begin{equation}
C_{\dot r n}(t)
=
\hat{\gamma}v_0
\int_0^\infty du\,\mu(u)\,e^{-|t-u|/\tau_{\rm a}}.
\label{eq:Crdotn_general_app}
\end{equation}
In particular,
\begin{equation}
C_{\dot r n}(0)
=
\hat{\gamma}v_0\int_0^\infty du\,\mu(u)e^{-u/\tau_{\rm a}}
=
\hat{\gamma}v_0\,\tilde\mu(\tau_{\rm a}^{-1}),
\end{equation}
where $\tilde\mu(\lambda)=\int_0^\infty dt\,e^{-\lambda t}\mu(t)$.

For the exponential kernel
\begin{equation}
\gamma(t)=\gamma_0\delta(t)+\frac{\gamma_1}{\tau_\mathrm{m}}e^{-t/\tau_\mathrm{m}}\Theta(t),
\qquad
\hat{\gamma}=\gamma_0+\gamma_1,
\end{equation}
the mobility kernel is
\begin{equation}
\mu(t)=\frac{1}{\gamma_0}\delta(t)-\frac{\gamma_1}{\gamma_0^2\tau_\mathrm{m}}e^{-t/\tau_{\rm v}}\Theta(t),
\qquad
\tau_{\rm v}=\frac{\gamma_0\tau_\mathrm{m}}{\gamma_0+\gamma_1}.
\label{eq:mu_explicit_app}
\end{equation}
Substituting Eq.~\ref{eq:mu_explicit_app} into Eq.~\ref{eq:Crdotn_general_app},
one obtains for $t\ge 0$
\begin{equation}
C_{\dot r n}(t)
=
(\gamma_0+\gamma_1)v_0
\left[
\frac{1}{\gamma_0}e^{-t/\tau_{\rm a}}
-\frac{\gamma_1}{\gamma_0^2\tau_\mathrm{m}}
\int_0^\infty du\,e^{-u/\tau_{\rm v}}e^{-|t-u|/\tau_{\rm a}}
\right].
\end{equation}
For $t\ge 0$ the integral splits at $u=t$,
\begin{equation}
\int_0^\infty du\,e^{-u/\tau_{\rm v}}e^{-|t-u|/\tau_{\rm a}}
=
\underbrace{\int_0^t e^{-u/\tau_{\rm v}}e^{-(t-u)/\tau_{\rm a}}\,du}_{I_1}
+
\underbrace{\int_t^\infty e^{-u/\tau_{\rm v}}e^{-(u-t)/\tau_{\rm a}}\,du}_{I_2},
\end{equation}
with
\begin{align}
I_1 &= \frac{\tau_{\rm a}\tau_{\rm v}}{\tau_{\rm a}-\tau_{\rm v}}
\left(e^{-t/\tau_{\rm a}}-e^{-t/\tau_{\rm v}}\right) \qquad
I_2 = \frac{\tau_{\rm a}\tau_{\rm v}}{\tau_{\rm a}+\tau_{\rm v}}\,e^{-t/\tau_{\rm v}}.
\end{align}
Combining over the common denominator $\tau_{\rm a}^2-\tau_{\rm v}^2$ gives
\begin{equation}
\int_0^\infty du\,e^{-u/\tau_{\rm v}}e^{-|t-u|/\tau_{\rm a}}
=
\frac{\tau_{\rm a}\tau_{\rm v}}{\tau_{\rm a}^2-\tau_{\rm v}^2}
\Big[
(\tau_{\rm a}+\tau_{\rm v})\,e^{-t/\tau_{\rm a}}
-2\tau_{\rm v}\,e^{-t/\tau_{\rm v}}
\Big],
\end{equation}
so that
\begin{equation}
C_{\dot r n}(t)
=
(\gamma_0+\gamma_1)v_0
\left[
\frac{1}{\gamma_0}e^{-t/\tau_{\rm a}}
-\frac{\gamma_1}{\gamma_0^2\tau_\mathrm{m}}
\frac{\tau_{\rm a}\tau_{\rm v}}{\tau_{\rm a}^2-\tau_{\rm v}^2}
\Big(
(\tau_{\rm a}+\tau_{\rm v})\,e^{-t/\tau_{\rm a}}
-2\tau_{\rm v}\,e^{-t/\tau_{\rm v}}
\Big)
\right],
\quad t\ge 0,
\label{eq:Crdotn_explicit_app}
\end{equation}
for $\tau_{\rm a}\neq \tau_{\rm v}$.
It is convenient to rewrite this result as
\begin{equation}
C_{\dot r n}(t)
=
v_0\Big[
A\,e^{-t/\tau_{\rm a}}
+
B\,e^{-t/\tau_{\rm v}}
\Big],
\qquad t\ge 0,
\label{eq:Crdotn_AB_form}
\end{equation}
with
\begin{equation}\label{eq:An_Bn_def}
A
=
(\gamma_0+\gamma_1)
\left[
\frac{1}{\gamma_0}
-
\frac{\gamma_1\tau_{\rm a}\tau_{\rm v}}{\gamma_0^2\tau_\mathrm{m}(\tau_{\rm a}-\tau_{\rm v})}
\right] \qquad
B
=
(\gamma_0+\gamma_1)
\frac{2\gamma_1\tau_{\rm a}\tau_{\rm v}^2}{\gamma_0^2\tau_\mathrm{m}(\tau_{\rm a}^2-\tau_{\rm v}^2)}.
\end{equation}
Thus, also in this case the correlation is a superposition of two relaxation modes,
one controlled by the orientational persistence time $\tau_{\rm a}$ and one by the
emergent relaxation time $\tau_{\rm v}$.

At the degenerate point $\tau_{\rm a}=\tau_{\rm v}$,
Eq.~\ref{eq:Crdotn_explicit_app} must be replaced by its smooth limit.
Expanding $I_1+I_2$ to first order in $\tau_{\rm v}-\tau_{\rm a}$ and taking the limit gives
\begin{equation}
C_{\dot r n}(t)
=
(\gamma_0+\gamma_1)v_0
\left[
\frac{1}{\gamma_0}e^{-t/\tau_{\rm a}}
-
\frac{\gamma_1}{\gamma_0^2\tau_\mathrm{m}}
\left(t+\frac{\tau_{\rm a}}{2}\right)e^{-t/\tau_{\rm a}}
\right],
\qquad t\ge 0.
\label{eq:Crdotn_degenerate_app}
\end{equation}
Evaluating Eq.~\ref{eq:Crdotn_explicit_app} at $t=0$ gives
\begin{equation}
C_{\dot r n}(0)
=
v_0\left(
1+
\frac{\gamma_1\tau_\mathrm{m}}
{\gamma_0(\tau_{\rm a}+\tau_\mathrm{m})}
\right),
\label{eq:Crdotn_t0_app}
\end{equation}
which shows that the instantaneous velocity-orientation correlation is enhanced
with respect to the Markovian value $v_0$ by the viscoelastic contribution.

\subsubsection{Effective persistence}\label{sec:tau_eff_calculation}

Here we derive the effective persistence and run-length scales introduced in the main text from the velocity--orientation correlation function of an isolated active particle. We show that the competition between the active persistence time $\tau_{\rm a}$ and the viscoelastic memory time $\tau_{\rm m}$ generates a non-monotonic effective persistence, providing a microscopic interpretation of the suppression and recovery of MIPS in terms of a renormalized effective run length.

The re-entrant behavior originates from the competition between active persistence and delayed viscoelastic relaxation. Although the long-time friction is fixed by $\hat{\gamma}=\gamma_0+\gamma_1$, the instantaneous mobility remains frequency dependent,
\begin{equation}
\hat{\gamma}(\omega)
=
\gamma_0+\frac{\gamma_1}{1+i\omega\tau_{\rm m}}.
\end{equation}
As a consequence, the active propulsion cannot be characterized solely by the bare velocity $v_0$, but rather by an effective, history-dependent propulsion strength. For $\tau_{\rm m}\sim\tau_{\rm a}$, the environment stores elastic stress over a time comparable to the orientational persistence, so that delayed viscoelastic response opposes the newly reoriented propulsion direction and reduces directional persistence. This weakens the self-trapping mechanism responsible for MIPS and explains the fluidization observed at intermediate memory times. By contrast, for $\tau_{\rm m}\gg\tau_{\rm a}$, the bath relaxes much more slowly than the active orientation, and the short-time dynamics becomes dominated by the reduced instantaneous friction $\gamma_0$, effectively enhancing propulsion and restoring phase separation.

This picture can be reframed in terms of a renormalized effective run length $\ell_{\rm eff}$, which controls MIPS. To this end, we consider the velocity-orientation correlation function $C_{\dot{r}n}(t)$ \eqref{eq:C_rdotn_def} from which one may define both an effective persistence time
\begin{equation}
\tau_{\rm eff} = \int_0^\infty dt \abs{\frac{C_{\dot{r}n}(t)}{C_{\dot{r}n}(0)}}  ,
\end{equation}
and an effective run length
\begin{equation}
\ell_{\rm eff} = \int_0^\infty dt \abs{C_{\dot{r}n}(t)} .
\end{equation}
Furthermore, one has $\ell_{\rm eff} \approx v_{\rm eff}\tau_{\rm eff}$, with $v_{\rm eff} = \langle \dot{\pmb{r}}\cdot\pmb{n} \rangle$. The non-monotonic dependence of $\tau_{\rm eff}$ and $\ell_{\rm eff}$ on $\tau_\mathrm{m}$ provides direct evidence that viscoelastic memory renormalizes the effective active run length, thereby controlling the onset, suppression, and reappearance of MIPS. 

\begin{figure}[t!]
    \centering    \includegraphics[width=\linewidth]{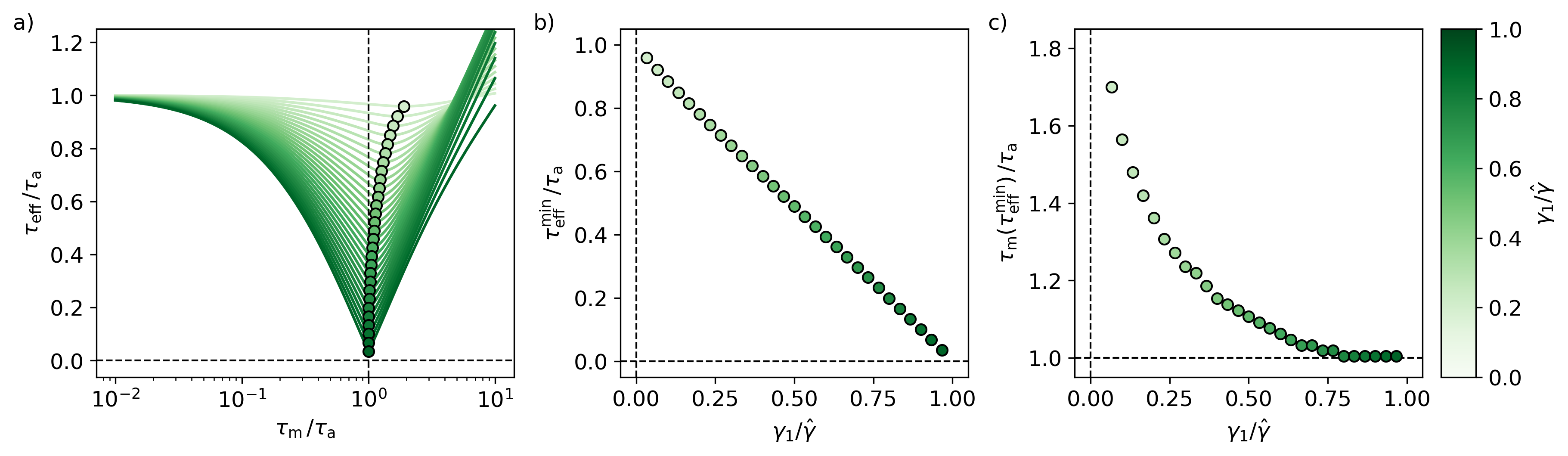}
    \caption{
Dependence of the effective persistence time on the viscoelastic coupling strength $\gamma_1/\hat{\gamma}$. 
a) Effective persistence time $\tau_{\rm eff}$ as a function of $\tau_{\rm m}/\tau_{\rm a}$ for different values of $\gamma_1/\hat{\gamma}$. All curves display a pronounced minimum near $\tau_{\rm m}\sim\tau_{\rm a}$, corresponding to the regime where delayed viscoelastic response most strongly suppresses directional persistence. 
b) Minimum value of the effective persistence time, $\tau_{\rm eff}^{\rm min}$, decreasing monotonically as a function of $\gamma_1/\hat{\gamma}$. 
c) Memory timescale $\tau_{\rm m}(\tau_{\rm eff}^{\rm min})$ at which the minimum of $\tau_{\rm eff}$ is reached. The position of the minimum converges to $\tau_{\rm a}$ as $\gamma_1/\hat{\gamma}$ increases. Colors indicate the value of $\gamma_1/\hat{\gamma}$.
}
    \label{fig:tau_eff_min}
\end{figure}

In Fig.~\ref{fig:tau_eff_min}, we characterize how the suppression of persistence depends on the relative strength of the viscoelastic coupling $\gamma_1/\hat{\gamma}$. As shown in Fig.~\ref{fig:tau_eff_min}a, all curves display a pronounced non-monotonic dependence on the memory timescale, with a minimum occurring for $\tau_{\rm m}\sim\tau_{\rm a}$. Increasing the viscoelastic contribution deepens this minimum, indicating that delayed environmental response progressively enhances the anti-persistent effects of the dynamics. The minimum value of the effective persistence time decreases monotonically with $\gamma_1/\hat{\gamma}$, as shown in Fig.~\ref{fig:tau_eff_min}b. At the same time, Fig.~\ref{fig:tau_eff_min}c shows that the position of the minimum tends to $\tau_{\rm a}$ as the viscoelastic contribution, i.e. $\gamma_1/\hat{\gamma}$, increases, remaining of the order of the active persistence time over the full range of viscoelastic couplings explored. These results confirm that the onset of anti-persistent dynamics is governed primarily by the competition between the environmental memory time and the orientational persistence time, while the magnitude of the suppression is controlled by the strength of the viscoelastic response.

\subsection{Markovian embedding}\label{sec:Mark_mem}

In this section we show how the exponential memory kernel used in the main text can be obtained by integrating out an auxiliary environmental degree of freedom. We consider the coupled overdamped dynamics
\begin{equation}
\gamma_0 \dot{\pmb{r}}^{\,i} (t)
=
\pmb{F}(\pmb{r}^{\,i})
+
\hat{\gamma} \pmb{v}^{\,i}(t)
-
\frac{\partial U_{\rm int}(\pmb{r}^{\,i},\pmb{r}_m^{\,i})}{\partial \pmb{r}^{\,i}}
+
\pmb{\xi}_r^{\,i},
\label{eq:x_markovian_embedding}
\end{equation}
\begin{equation}
\gamma_1 \dot{\pmb{r}}_m^{\,i} (t)
=
-
\frac{\partial U_{\rm int}(\pmb{r}^{\,i},\pmb{r}_m^{\,i})}{\partial \pmb{r}_m^{\,i}}
+
\pmb{\xi}_m^{\,i},
\label{eq:y_markovian_embedding}
\end{equation}
where $i$ labels different particles, $\pmb{r}^{\,i}$ denotes the physical particle coordinate, and $\pmb{r}_m^{\,i}$ an auxiliary environmental degree of freedom interacting via a harmonic potential
\begin{equation}
U_{\rm int}(\pmb{r},\pmb{r}_m)
=
\frac{k}{2}
\left(
\pmb{r}-\pmb{r}_m
\right)^2 .
\end{equation}

The noises are Gaussian and white,
$\left\langle
\xi_{\mu,\alpha}^{\,i}(t)
\xi_{\nu,\beta}^{\,j}(t')
\right\rangle
=
2k_{\rm B}T\gamma_\mu\,
\delta_{ij}
\delta_{\alpha\beta}
\delta_{\mu\nu}
\delta(t-t')$, where $\mu,\nu\in\{r,m\}$ denote the two coupled degrees of freedom, while $\alpha,\beta$ label Cartesian components. The equation for the auxiliary variable becomes
\begin{equation}
\gamma_1 \dot{\pmb{r}}_m^{\,i}
=
-k(\pmb{r}_m^{\,i}-\pmb{r}^{\,i})
+
\pmb{\xi}_m^{\,i}.
\label{eq:y_dynamics_explicit}
\end{equation}
Introducing $\tau_\mathrm{m}=\gamma_1/k$ and taking the Laplace transforms
one obtains
\begin{equation}
\left[
s\,\hat{\pmb{r}}_m^{\,i}(s)-\pmb{r}_m^{\,i}(0)
\right]
=
-\tau_\mathrm{m}^{-1}\hat{\pmb{r}}_m^{\,i}(s)
+
\tau_\mathrm{m}^{-1}\hat{\pmb{r}}^{\,i}(s)
+
{\gamma^{-1}_m} \hat{\pmb{\xi}}_m^{\,i}(s).
\end{equation}
Solving for $\hat{\pmb{r}}_m^{\,i}(s)$ gives
\begin{equation}
\hat{\pmb{r}}_m^{\,i}(s)
=
\frac{1}{s+\tau_\mathrm{m}^{-1}} \left[ \pmb{r}_m^{\,i}(0) + \tau_\mathrm{m}^{-1} \hat{\pmb{r}}^{\,i}(s) + {\gamma_1^{-1}} \hat{\pmb{\xi}}_m^{\,i}(s) \right]
\end{equation}
Taking the inverse Laplace transform,
\begin{equation}
\pmb{r}_m^{\,i}(t)
=
\pmb{r}_m^{\,i}(0)e^{-t/\tau_\mathrm{m}}
+
\tau_\mathrm{m}^{-1}
\int_0^t ds\,
e^{-(t-s)/\tau_\mathrm{m}}\pmb{r}^{\,i}(s)
+
\gamma_1^{-1}
\int_0^t ds\,
e^{-(t-s)/\tau_\mathrm{m}}\pmb{\xi}_m^{\,i}(s).
\label{eq:y_solution_real_time}
\end{equation}
For notational simplicity, we now drop the particle index $i$. Substituting Eq.~\ref{eq:y_solution_real_time} into Eq.~\ref{eq:x_markovian_embedding} gives
\begin{align}
\gamma_0 \dot{\pmb{r}}(t)
=
\pmb{F}(\pmb{r})
+
\hat{\gamma}\pmb{v}(t)
-k\pmb{r}(t)
+
k\pmb{r}_m(0)e^{-t/\tau_\mathrm{m}}
+
\tau_\mathrm{m}^{-1} k
\int_0^t ds\,
e^{-(t-s)/\tau_\mathrm{m}}\pmb{r}(s)
+
\pmb{\eta}(t),
\label{eq:x_after_substitution}
\end{align}
where
\begin{equation}
\pmb{\eta}(t)
=
\pmb{\xi}_r(t)
+
\frac{k}{\gamma_1}
\int_0^t ds\,
e^{-(t-s)/\tau_\mathrm{m}}\pmb{\xi}_m(s).
\end{equation}
Integrating the convolution term by parts,
\begin{align}
\int_0^t ds\,
e^{-(t-s)/\tau_\mathrm{m}}\pmb{r}(s)
=
\tau_\mathrm{m} \pmb{r}(t)
-
\tau_\mathrm{m} \pmb{r}(0)e^{-t/\tau_\mathrm{m}}
-
\tau_\mathrm{m}
\int_0^t ds\,
e^{-(t-s)/\tau_\mathrm{m}}\dot{\pmb{r}}(s).
\end{align}
Using $\tau_\mathrm{m}=\gamma_1/k$, the instantaneous terms proportional to $\pmb{r}(t)$ cancel exactly, yielding
\begin{align}
\int_0^t ds\,
\gamma(t-s)
\dot{\pmb{r}}(s)
=
\pmb{F}(\pmb{r})
+
\hat{\gamma}\pmb{v}(t)
+
\pmb{\eta}(t)
+
k\left[
\pmb{r}_m(0)-\pmb{r}(0)
\right]
e^{-t/\tau_\mathrm{m}}.
\label{eq:GLE_final_embedding}
\end{align}
This is a generalized Langevin equation with exponentially decaying transient term depending on the initial condition which vanishes in the steady state and
\begin{equation}
\gamma(t)
=
\gamma_0\delta(t)
+
\frac{\gamma_1}{\tau_\mathrm{m}}
e^{-t/\tau_\mathrm{m}}\Theta(t),
\qquad \langle \eta^{i}(t)\eta^{j}(s)\rangle = k_{\rm B} T \gamma(|t-s|)
\end{equation}

\section{Simulation methods}
\label{sec:simulation_methods}

In this section, we describe the numerical implementation of the model and
the observables used to characterize the collective dynamics. We first detail
the simulation protocol, parameter choices, and numerical methods employed in
the integration of the dynamics (Sec.~\ref{sec:supp_simulation_details}). We then introduce the quantities used to
analyze phase behavior and clustering properties in the simulated systems (Sec.~\ref{sec:observables}).

\subsection{Simulation details and parameter choices}
\label{sec:supp_simulation_details}

We perform numerical simulations of active Brownian particles with a memory
kernel in two spatial dimensions using periodic boundary conditions and the markovian embedding described in Sec.~\ref{sec:Mark_mem} (Eqs.~\ref{eq:x_markovian_embedding} and \ref{eq:y_markovian_embedding}). The two
control parameters varied throughout the paper are the Péclet number,
\(\mathrm{Pe}\), and the ratio between the memory and active persistence
timescales, \(\tau_\mathrm{m}/\tau_\mathrm{a}\). All other parameters are kept fixed unless
stated otherwise.

The system contains \(N\) particles at packing fraction \(\phi\). The packing
fraction is related to the particle diameter \(\sigma\) through
\begin{equation}
    \phi = \frac{N \pi \sigma^2}{4V},
\end{equation}
where \(V\) is the area of the simulation box.

Particles interact via a purely repulsive soft potential. The corresponding
pairwise interaction force between particles \(i\) and \(j\) separated by a
distance \(r_{ij}\) is
\begin{equation}
    \pmb{F}_{ij} =
    \begin{cases}
    -\dfrac{\epsilon}{\sigma}
    \left(1 - \dfrac{r_{ij}}{\sigma}\right)
    \hat{\pmb{r}}_{ij},
    & r_{ij} < \sigma, \\[8pt]
    0,
    & r_{ij} \geq \sigma,
    \end{cases}
\end{equation}
where \(\hat{\pmb{r}}_{ij}\) is the unit vector joining the particle
centers, \(\sigma\) is the particle diameter, and \(\epsilon\) sets the
interaction energy scale. Equivalently, the interaction stiffness may be
written as
\begin{equation}
    \epsilon = \kappa \sigma^2 .
\end{equation}
The value of \(\epsilon\) is chosen such that the typical equilibrium overlap
between interacting particles is \(\delta\).

The positional degree of freedom is coupled to a friction coefficient
\(\gamma_0\), while the auxiliary memory degree of freedom is coupled to a
friction coefficient \(\gamma_1\). We define the total friction coefficient as
\begin{equation}
    \hat{\gamma} = \gamma_0 + \gamma_1 .
\end{equation}

Thermal fluctuations are introduced through diffusion coefficients associated
with both the positional and auxiliary memory degrees of freedom, as stated in Sec.~\ref{sec:Mark_mem}. We set the same thermal energy scale $k_{\mathrm{B}}T$ for both degrees of freedom. This choice
ensures consistency between dissipation and thermal noise in the passive
limit. In practice, the results presented here use negligible temperature $k_BT=10^{-12}$.

The unit of time is set by the elastic relaxation timescale
\begin{equation}
    \tau_{\mathrm{el}} = \frac{\gamma_0}{\kappa},
\end{equation}
which corresponds to the relaxation timescale associated with the soft
interaction force. Simulations are integrated with timestep
\begin{equation}
    \Delta t = 0.1\,\tau_{\mathrm{el}}.
\end{equation}
Each trajectory is evolved for a total simulation time
\begin{equation}
    T_{\mathrm{sim}} = 200\,\max(\tau_\mathrm{m},\tau_\mathrm{a}),
\end{equation}
which is much larger than the active persistence timescale.

To accelerate the computation of pair interactions, we employ a Verlet
neighbor list combined with a cell-list decomposition of the simulation box.
The cell list is used to efficiently identify nearby particles, while the
Verlet list stores neighboring pairs within an enlarged cutoff radius.
Neighbor lists are updated periodically according to the particle
displacements.

Unless otherwise stated, simulations are performed at fixed packing fraction
\(\phi = 0.4\), particle number \(N = 5000\), overlap parameter
\(\delta = 0.05\), friction coefficients
\(\gamma_0 = 0.1\), \(\gamma_1 = 0.9\), \(\hat \gamma = 1\) and temperature \(k_BT = 10^{-12}\). The control parameters varied
throughout the paper are the Péclet number \(\mathrm{Pe}\) and the ratio
\(\tau_\mathrm{m}/\tau_\mathrm{a}\).
The comparison with standard active Brownian particles in the absence of
memory (leftmost columns of the phase diagrams in
Fig.~\ref{Fig2}b and Fig.~\ref{fig:dynamics}c is obtained in the limit
\(\gamma_1 = 0\), with \(\gamma_0 = 1\).

Initial conditions are chosen depending on the protocol investigated. In the first protocol (referred to as \emph{Cluster initialization} and used in Fig.~\ref{Fig2}), particles are initialized on a compact square lattice with nearest-neighbor distance equal to the particle diameter \(\sigma\). This configuration forms an initially dense cluster surrounded by an empty region in the remainder of the simulation box.  In the second protocol (referred to as \emph{Gas initialization} and used in Fig.~\ref{fig:dynamics}), particles are initialized on a square lattice distributed uniformly throughout the simulation box, resulting in a spatially homogeneous configuration at the initial time. These two preparation protocols allow us to probe both homogeneous and
phase-separated initial states. 

\subsection{Observables and clustering analysis}
\label{sec:observables}

In this subsection, we describe the observables used to characterize the
collective behavior and phase separation properties of the system. In
particular, we focus on the cluster fraction used throughout the paper to
quantify aggregation and motility-induced phase separation.

Clusters are identified using a contact criterion: two particles belong to the
same cluster when their separation is smaller than $0.99 \times \sigma$, with $\sigma$ the particle diameter. From this connectivity graph, we determine the size of the largest
cluster and define the cluster fraction
\begin{equation}
    c = \frac{N_{\mathrm{largest}}}{N},
\end{equation}
where \(N_{\mathrm{largest}}\) is the number of particles belonging to the
largest connected cluster and \(N\) is the total number of particles. The
quantity \(c\) is used throughout the paper as an order parameter for
motility-induced phase separation. Throughout the paper, $c$ is averaged over three distinct realizations. 

\begin{figure}
    \centering
    \includegraphics[width=0.7
    \linewidth]{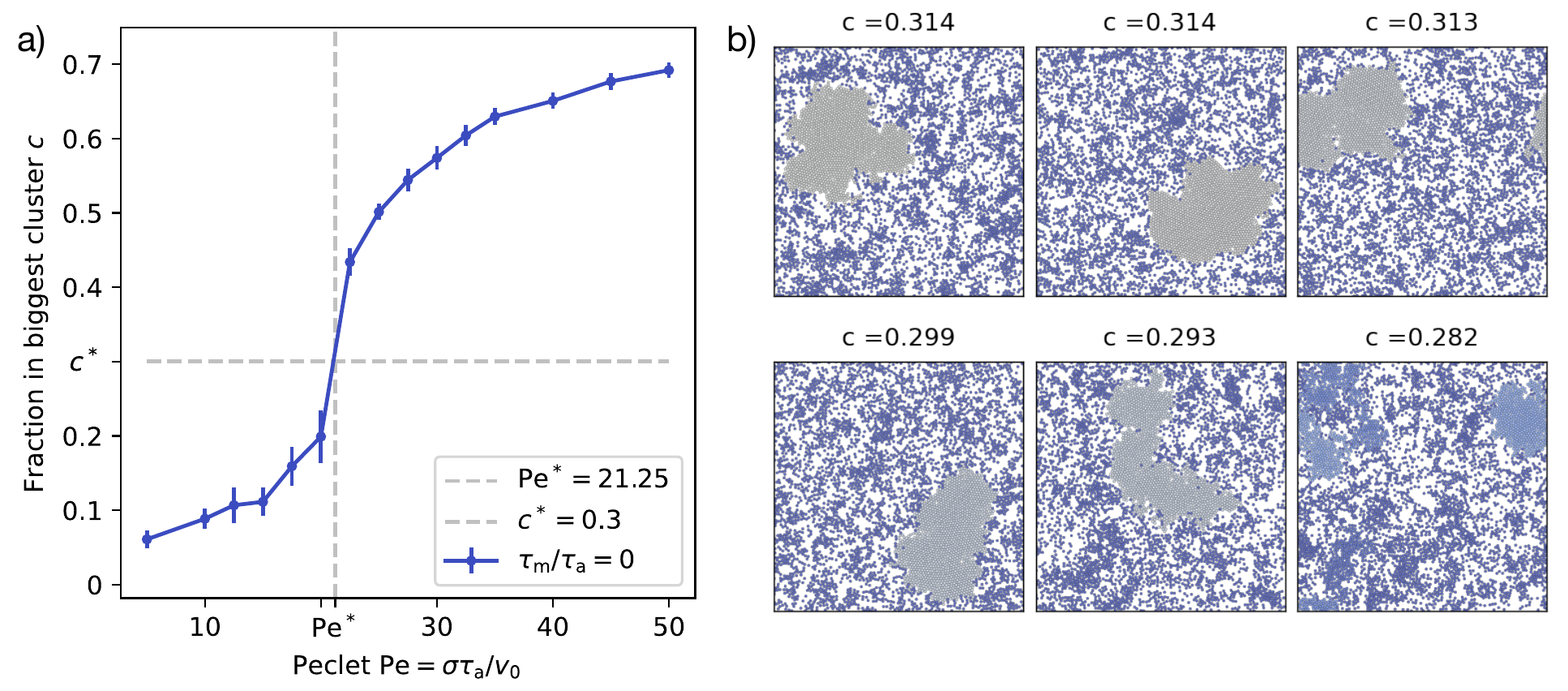}
    \caption{The fraction of particles in the largest cluster, $c$, provides a convenient order parameter for Motility-induced phase separation. a) In the absence of memory, the onset of phase separation occurs at  $\mathrm{Pe}^*=21.25$ corresponding to $c^*=0.3$. b) Simulation snapshots with memory for values of c near the transition confirm that $c^*$ provides a robust criterion for identifying phase separation.} 
    \label{fig:sup_cluster_threshold}
\end{figure}

To determine a threshold value for phase separation, we first calibrate the
observable in the standard active Brownian particle limit without memory
(\(\gamma_1 = 0\)). As shown in
Fig.~\ref{fig:sup_cluster_threshold}a, the onset of motility-induced phase
separation occurs at \(\mathrm{Pe}^* \simeq 21.25\), corresponding to a cluster
fraction \(c^* = 0.3\). We therefore use \(c^*=0.3\) as the criterion for the
presence of phase separation throughout the paper, which sets the color scale
used in Figs.\ref{Fig2}a-b and \ref{fig:dynamics}a of the main text. 

To verify that this criterion remains robust in the presence of memory,
we inspect representative simulation snapshots across our dataset.
As illustrated in Fig.~\ref{fig:sup_cluster_threshold}b, configurations with
\(c \lesssim 0.3\) do not exhibit stable dense clusters, while configurations with
\(c \gtrsim 0.3 \) display clear phase-separated structures.

\section{Collective dynamics}\label{sec:coll_dyn_supp}

In this section, we provide additional analytical and numerical results for the interacting system and the collective dynamics near the transition. We first show the convergence of the phase diagram boundary at large memory regimes (Sec.~\ref{sec:large_deborah}). 
We also analyze the interaction-induced corrections to the velocity-orientation correlation and compare the resulting effective velocity with numerical simulations (Sec.~\ref{sec:CnF_calculations}). 
We then present additional dynamical trajectories and transition-time measurements characterizing the slow relaxation and competing pathways observed near the onset of phase separation (Sec.~\ref{sec:supp_dynamics}). Finally, we show that changing the ratio of friction between the memory and position degree of freedom does not qualitatively affect the re-entrant MIPS (Sec.~\ref{sec:supp_fric_ratios}).

\subsection{Large-memory regime}
\label{sec:large_deborah}

Here, we characterize the asymptotic behavior of the system in the large-memory regime $\tau_\mathrm{m}/\tau_\mathrm{a}\gg1$. In this limit, the viscoelastic relaxation becomes much slower than the active persistence time, and the short-time dynamics is effectively governed by the instantaneous friction coefficient $\gamma_0$. The system therefore approaches the behavior of a Markovian active Brownian fluid with friction $\gamma_0$. By contrast, in the limit $\tau_\mathrm{m}=0$, the dynamics reduces to a Markovian system with total friction $\gamma_0+\gamma_1$.

As a consequence, the effective Péclet number in the large-memory regime is enhanced by a factor $(\gamma_0+\gamma_1)/\gamma_0$ compared with the Markovian reference system. The phase-separation threshold is therefore expected to approach
\begin{equation}
\mathrm{Pe}
=
\frac{\gamma_0}{\gamma_0+\gamma_1}
\mathrm{Pe}^*,
\end{equation}
where $\mathrm{Pe}^*$ denotes the critical Péclet number of the Markovian system shown in Fig.~\ref{fig:sup_cluster_threshold}a.

\begin{figure}
    \centering
    \includegraphics[width=0.4\linewidth]{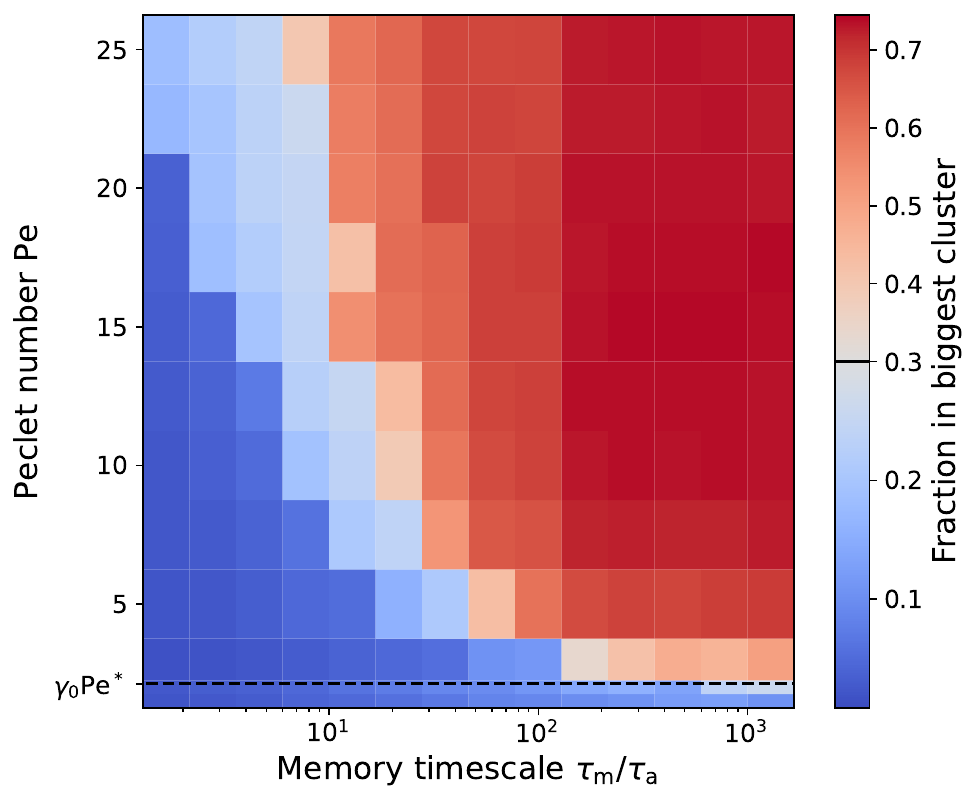}
    \caption{Phase-separation boundary in the large-memory regime. For $\tau_\mathrm{m}/\tau_\mathrm{a}\gg1$, the transition line saturates toward the effective Markovian prediction controlled by the instantaneous friction coefficient $\gamma_0$ (for $\gamma_0+\gamma_1=1)$. Simulations are initialized in a homogeneous state.}
    \label{fig:large_deborah}
\end{figure}

Figure~\ref{fig:large_deborah} confirms this prediction and shows that the phase boundary saturates toward a finite value of the Péclet number for large memory times.

\subsection{Interaction-induced renormalization of the effective velocity}
\label{sec:CnF_calculations}

Here, we compute the correction to the equal-time velocity-orientation correlation due to multi-particle interactions and in terms of the force-orientation correlation function $C_{nF}(t)$. Numerical measurements of $C_{nF}(0)$ for different activities, packing fractions, and viscoelastic couplings are then used to quantify the resulting correction of the effective propulsion velocity.

\begin{figure}[t!]
    \centering    \includegraphics[width=\linewidth]{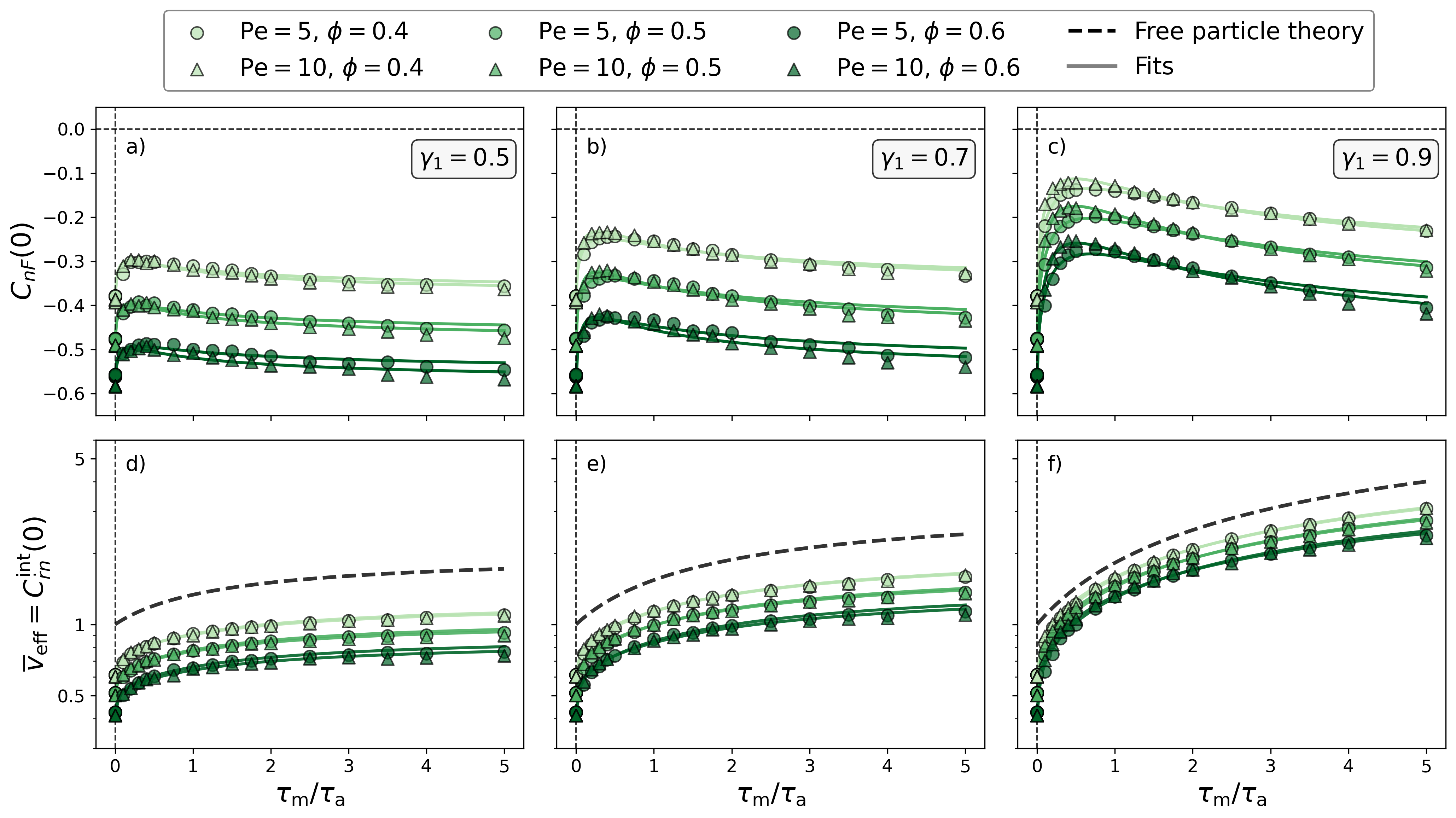}
    \caption{Figure showing the dependence of the force-orientation correlation
$C_{nF}(0)$ and the effective propulsion response
$\bar{v}_{\rm eff}=C_{\dot r n}^{\rm \, int}(0)$ on the memory timescale ratio
$\tau_\mathrm{m}/\tau_\mathrm{a}$ for different values of the viscoelastic coupling
parameter $\gamma_1$ calculated for $N=500$ particles.
Top panels (a-c): $C_{nF}(0)$ for different
combinations of activity $Pe$ and packing fractions $\phi$,
together with fits to the phenomenological form
$a+b\sqrt{\tau_\mathrm{m}/\tau_\mathrm{a}}$.
Bottom panels (d-f): corresponding estimates for the effective velocity $\bar{v}_{\rm eff} = C_{\dot r n}^{\rm \, int}(0)$ from simulations (scatter plot) compared to Eq. \ref{eq:C_nf_final} where $C_{nF}(0)$ is extrapolated using fitted curves (solid colored lines).
The dashed black line represents the free-theory effective velocity $C^{\rm \, free}_{\dot{r}n}(0)$.}
    \label{fig:CnF0_fits}
\end{figure}

Using the mobility representation of Eq. \ref{eq:GLE_general} 
\begin{equation}
\dot{\pmb r}^i(t)
=
\int_{-\infty}^t ds \,\mu(t-s)\,\left(\pmb F_{\mathrm{int}}^i(s)
+
\hat{\gamma}\,\pmb v_{\rm a}^i(s)
+
\boldsymbol{\eta}^i(s)\right),
\end{equation}
with mobility kernel
\begin{equation}
\mu(t)=\frac{1}{\gamma_0}\delta(t)
-
\frac{\gamma_1}{\gamma_0^2\tau_\mathrm{m}}
e^{-t/\tau_{\rm v}}\Theta(t),
\qquad
\tau_{\rm v}=\frac{\gamma_0}{\gamma_0+\gamma_1}\tau_\mathrm{m}\,,
\end{equation}
interaction forces given by 
combined \ref{eq:interaction} and exponentially correlated active propulsion (Eq.~\ref{eq:v_exp}), the equal-time velocity-orientation correlation can be written as
\begin{equation} \label{eq:Cvn_int}
C_{\dot{r}n}^{\rm \, int}(0)
=
C_{\dot{r}n}^{\rm \, free}(0)
+
\int_0^\infty ds\,\mu(s)\,C_{nF}(s),
\end{equation}
where
\begin{equation}
C_{nF}(s)
=
\left\langle
\pmb n_s\cdot \pmb F_{0}
\right\rangle \,.
\end{equation}
The force-orientation correlation inherits the same relaxation time as the orientational persistence. Indeed, as suggested by Eq. \ref{eq:ABP_iso_corr}, the first angular mode relaxes exponentially and obeys the following equation:
\begin{equation}
\frac{d}{dt} n^{(1)}_\alpha(t)
=
-D_\theta\langle n^{(1)}_\alpha(t)\rangle.
\end{equation}
This first moment is the only one contributing to correlations with other observables. In particular, multiplying by $F_\alpha(0)$ (and dropping the $(1)$ superscript) and averaging yields
\begin{equation}
\frac{d}{dt}
\langle n_\alpha(t)F_\alpha(0)\rangle
=
-D_\theta
\langle n_\alpha(t)F_\alpha(0)\rangle,
\end{equation}
since the rotational noise is uncorrelated with the force evaluated at the initial time. Summing over Cartesian components gives
\begin{equation}
\frac{d}{dt}C_{nF}(t)
=
-\frac{1}{\tau_\mathrm{a}}C_{nF}(t),
\end{equation}
whose solution is
\begin{equation}
C_{nF}(t)=C_{nF}(0)e^{-t/\tau_\mathrm{a}}.
\end{equation}
This exponential decay with typical time $\tau_{\rm a}$ has been verified numerically.

The interaction contribution can then be evaluated explicitly,
\begin{equation}\label{eq:C_nf_final}
\int_0^\infty ds\,\mu(s)\,C_{nF}(s)
=
\frac{C_{nF}(0)}{\gamma_0}
\left(
1-
\frac{\gamma_1\tau_\mathrm{a}}
{\gamma_0\tau_\mathrm{m}+\left(\gamma_0+\gamma_1\right)\tau_\mathrm{a}}
\right).
\end{equation}
Therefore,
\begin{equation}\label{eq:Cvn_int_fin}
C_{\dot{r}n}^{\rm \, int}(0)
=
C_{\dot{r}n}^{\rm \, free}(0)
+
\frac{C_{nF}(0)}{\gamma_0}
\left(
1-
\frac{\gamma_1\tau_\mathrm{a}}
{\gamma_0\tau_\mathrm{m}+\left(\gamma_0+\gamma_1\right)\tau_\mathrm{a}}
\right).
\end{equation}
Since $C_{nF}(0)<0$ for repulsive interactions, collisions reduce the alignment between velocity and propulsion direction generated by the viscoelastic memory kernel. 

In Fig.~\ref{fig:CnF0_fits}, we show the value of $C_{nF}(0)$ calculated numerically for different Péclet number $\mathrm{Pe}$, memory timescale $\tau_\mathrm{m}/\tau_\mathrm{a}$, viscoelastic coupling parameter $\gamma_1$ and packing fraction $\phi$, in systems of $N=500$ particles. We find that the $\tau_{\rm m}$ dependence of $C_{nF}(0)$ exhibits a robust phenomenological form across the range of parameters explored,
\begin{equation}
C_{nF}(0)
\simeq
v_0\phi
\left(
-a
+
\frac{
b\sqrt{\tau_\mathrm{m}}
}{
1+\tau_\mathrm{m}/\tau_{nF}
}
\right),
\label{eq:CnF_fit}
\end{equation}
as shown in Fig.~\ref{fig:CnF0_fits} (a-c). We find that $a \approx \hat{\gamma} = \gamma_0 + \gamma_1$ and is of the same order of magnitude of $b$,  while $\tau_{nF}$ characterizes the saturation timescale of the interaction-force correlations and is of the same order of $\tau_{\rm a}$. The prefactor $\phi$ captures the increasing contribution of collisions at larger packing fractions, compatible with known results for Markovian dynamics. As a consequence, at larger $\tau_{\rm m}$, the correction saturates once the viscoelastic relaxation becomes slower than the characteristic interaction decorrelation time $\tau_{nF}$.

Substituting Eq.~\ref{eq:CnF_fit} into Eq.~\ref{eq:Cvn_int_fin} yields, in the regime $\tau_{\rm v}/\tau_{\rm a}\ll1$,
\begin{equation}
\bar{v}_{\rm eff}
\simeq
v_0(1-\phi)
\left(
1+b\sqrt{\tau_\mathrm{m}/\tau_\mathrm{a}}
\right)\,.
\label{eq:veff_scaling}
\end{equation}
This scaling captures the enhancement of the effective propulsion velocity observed numerically in Fig.~\ref{fig:CnF0_fits} (d-f), and differs qualitatively from the dilute single-particle prediction (after rescaling by $1-\phi$), which displays only a weak low-memory enhancement before saturating. 

\subsection{Slow dynamics near the phase-separation transition}
\label{sec:supp_dynamics}

In this section, we present additional stochastic trajectories and transition-time measurements characterizing the dynamical crossover region near the onset of phase separation. We show that the relaxation dynamics becomes strongly dependent on fluctuations and preparation protocol close to the transition, with coexistence of dilute, transient, and long-lived clustered trajectories. We further demonstrate that the longest transition times are concentrated near the phase boundary, reflecting the competition between delayed viscoelastic relaxation and active self-trapping.

\begin{figure}
    \centering
    \includegraphics[width=\linewidth]{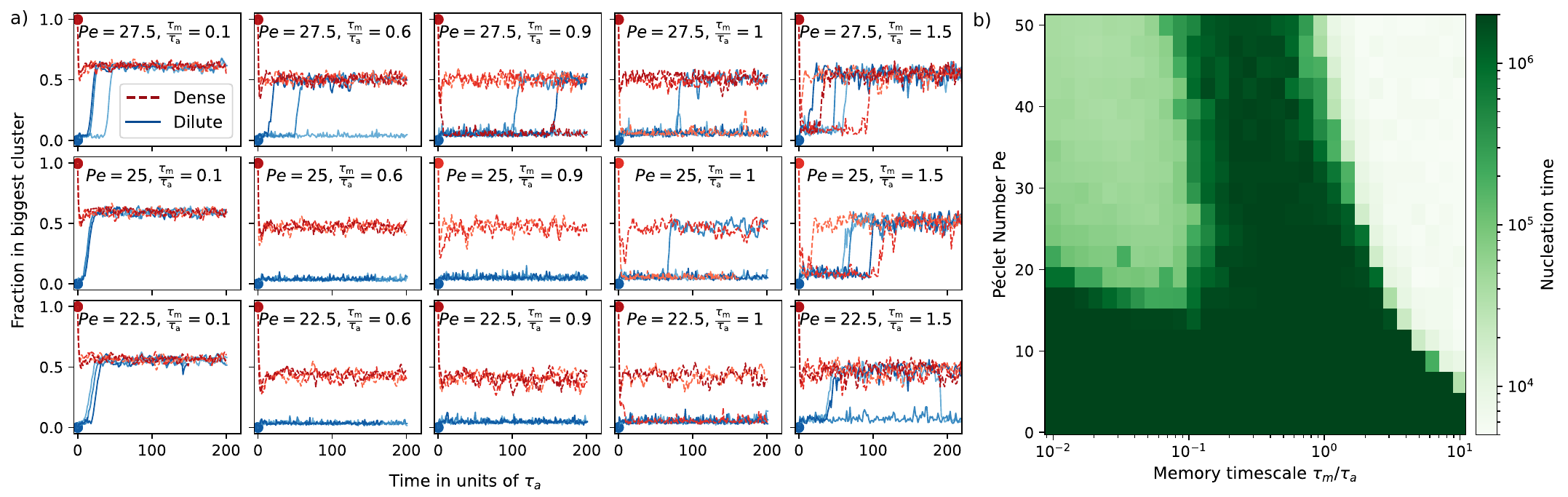}
    \caption{Slowdown near the transition. a) Additional stochastic trajectories of the cluster fraction for different parameters and initial conditions. Time is measured in units of $\tau_\mathrm{a}$, which varies with $\mathrm{Pe}$. b) Phase diagram of the transition time measured from homogeneous initial conditions, shown in elastic time units. Dark green regions where the nucleation time approaches $2\times 10^6$ indicate the absence of phase separation within the simulation time window.}
    \label{fig:supp_dynamics}
\end{figure}

Figure~\ref{fig:supp_dynamics}a shows additional stochastic trajectories of the cluster fraction for several points of the phase diagram, starting either from a homogeneous gas configuration (blue) or from an initially phase-separated state (red). Deep inside the homogeneous or phase-separated regions, both initial conditions converge toward the same long-time behavior, indicating the absence of metastability. This is the case, for example, for $\tau_\mathrm{m}/\tau_\mathrm{a}=0.1$ (leftmost column), or for $\tau_\mathrm{m}/\tau_\mathrm{a}=1.5$ and large $\mathrm{Pe}$, where all realizations phase separate.

Closer to the transition region, however, the long-time dynamics becomes strongly history dependent. For instance, at $\mathrm{Pe}=22.5$ and $\tau_\mathrm{m}/\tau_\mathrm{a}=0.6$, the initial condition almost entirely determines the final state: homogeneous initial conditions remain dilute whereas phase-separated initial conditions remain clustered. Intermediate situations are also observed, such as for $\tau_\mathrm{m}/\tau_\mathrm{a}=1$, where different stochastic realizations may evolve toward either state. In some cases, transient phase separation occurs before clusters eventually dissolve again, as illustrated for $\mathrm{Pe}=22.5$ and $\tau_\mathrm{m}/\tau_\mathrm{a}=1.5$. Altogether, these trajectories highlight the coexistence of competing relaxation pathways near the transition. The time axis in panel a is expressed in units of the active persistence time $\tau_\mathrm{a}$. Since $\tau_\mathrm{a}$ itself depends on $\mathrm{Pe}$, absolute times should not be directly compared between different lines.

Figure~\ref{fig:supp_dynamics}b shows the full phase diagram of the transition time obtained from homogeneous initial conditions, now represented in absolute units corresponding to the elastic timescale. White regions correspond to parameters for which phase separation does not occur within the simulation time window. Remarkably, the largest transition times are concentrated near the phase boundary separating homogeneous and phase-separated states. This pronounced slowdown reflects the competition between viscoelastic relaxation and active persistence close to the onset of self-trapping, where fluctuations strongly hinder the stabilization of dense clusters.

\subsection{Variation of the friction ratio}\label{sec:supp_fric_ratios}

\begin{figure}
    \centering
    \includegraphics[width=0.99\linewidth]{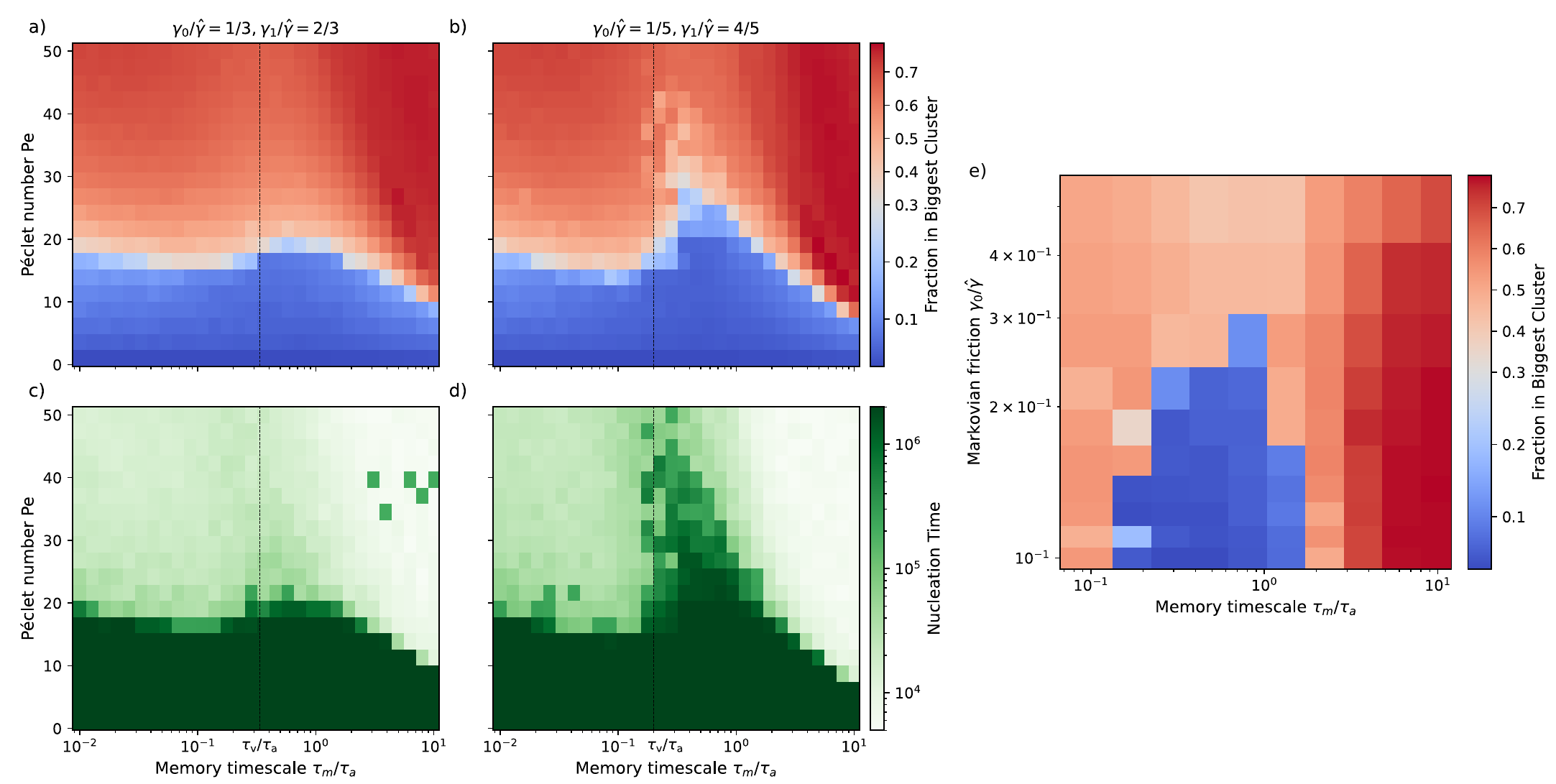}
    \caption{MIPS at different friction ratios for simulations initialized in a homogenized state. Top row (a, b): Fraction of particles in the system's largest cluster for $\gamma_0/\hat{\gamma} = 1/3,  \gamma_1/\hat{\gamma} = 2/3$ (panel a) and $\gamma_0/\hat{\gamma} = 1/5, \gamma_1/\hat{\gamma} = 4/5$ (panel b). Here, $\hat{\gamma} = \gamma_0 + \gamma_1$. Bottom row (c, d): Nucleation time in units of the elastic time scale for the same friction ratios shown in panels a and b. MIPS suppression can be seen in both rows near $\tau_{\rm m} \sim \tau_{\rm a}$, though at lower non-Markovian friction values $\gamma_1$, clusters are more stable with shorter nucleation times in this regime. e) Fraction of particles in the system's largest cluster for different values of the \emph{Markovian friction} $\gamma_0/\hat{\gamma}$ and the memory timescale ratio at the near-critical Péclet $\text{Pe} = 22.5$. Small values of $\gamma_0$ show a larger suppression of MIPS near $\tau_m = \tau_{\rm a}$, while larger Markovian frictions exhibit less suppression. Notice that, in all cases, only when $\tau_m \gg \tau_{\rm a}$ do we observe large clusters where the fraction of particles in the largest cluster $> 0.7$. }
    \label{fig:suppfig_fric_ratios}
\end{figure}

In this section, we present additional analyses of MIPS in systems with different values of the Markovian and non-Markovian frictions, $\gamma_0$ and $\gamma_1$, respectively, for systems initialized in a homogeneous state. In Sec.~\ref{sec:metastability} in the main text, we show that MIPS is suppressed near $\tau_{\rm m} \sim \tau_{\rm a}$ for $\gamma_0 = 1/10, \gamma_1 = 9/10$. In Fig.~\ref{fig:suppfig_fric_ratios}(a-d), we show that we see the same qualitative suppression of MIPS near $\tau_{\rm m} \sim \tau_{\rm a}$ for $\gamma_0 = 1/3, \gamma_1 = 2/3$ and $\gamma_0 = 1/5, \gamma_1 = 4/5$. For each system, $\hat{\gamma} = \gamma_0 + \gamma_1 = 1$, such that all systems in the $\tau_M \to 0$ limit behave as if in contact with a Markovian bath with friction coefficient $\hat{\gamma}$. We find that the nucleation time also diverges when MIPS is suppressed with these other friction coefficients, echoing the observation stated in Fig.~\ref{fig:dynamics} in the main text. In Fig.~\ref{fig:suppfig_fric_ratios}e, we further show that MIPS suppression at the near-threshold Péclet ${\rm Pe} = 22.5$ is maximized when $\gamma_0$ is small compared to $\gamma_1$. This indicates MIPS suppression is due to both the strength and timing of the medium's viscoelasticity.

\end{document}